\documentclass[preprint,5p,twocolumn]{elsarticle}

\usepackage{lineno,hyperref}
\modulolinenumbers[1]
\usepackage{amsmath}
\usepackage{siunitx}
\biboptions{numbers,sort&compress}

\usepackage{tikz} 
\usepackage[europeanresistors,americaninductors]{circuitikz}
\usepackage{adjustbox}
\usepackage{eurosym}
\usepackage{xspace}
\usepackage{caption}
\usepackage{booktabs}
\usepackage{tabularx}
\usepackage{threeparttable}
\usepackage{multicol}
\usepackage{float}
\usepackage{graphicx,dblfloatfix}
\usepackage{csvsimple}
\usepackage{url}
\usepackage{lipsum}
\usepackage{framed} %
\usepackage{nomencl} %

\makenomenclature

\graphicspath{{figures/}}

\journal{}

\newcommand{\ubar}[1]{\text{\b{$#1$}}}

\newcommand\picsize{1}

\def\co{CO${}_2$}
\def\el{${}_{\textrm{el}}$}
\def\th{${}_{\textrm{th}}$}
\def\h2{${}_{\textrm{H2}}$}

\begin{document}

\begin{frontmatter}
\title{Impact of climatic, technical and economic uncertainties on the optimal design of a coupled fossil-free electricity, heating and cooling system in Europe}
\author[aarhus]{K.~Zhu\corref{cor1}}
\ead{kunzhu@eng.au.dk}
\author[aarhus,iclimate]{M.~Victoria}
\author[aarhus,iclimate]{G. B.~Andresen}
\author[aarhus,iclimate]{M.~Greiner}
\cortext[cor1]{Corresponding author}
\address[aarhus]{Department of Engineering, Aarhus University, Aarhus, Denmark}
\address[iclimate]{iCLIMATE Interdisciplinary Centre for Climate Change, Aarhus University, Roskilde, Denmark}

\begin{abstract}
To limit the global temperature increase to 1.5~\si{\celsius}, fossil-free energy systems will be required eventually. To understand how such systems can be designed, the current state-of-the-art is to apply techno-economical optimisation modelling with high spatial and temporal resolution. This approach relies on a number of climatic, technical and economic predictions that reach multiple decades into the future. In this paper, we investigate how the design of a fossil-free energy system for Europe is affected by changes in these assumptions. In particular, the synergy among renewable generators, power-to-heat converters, storage units, synthetic gas and transmission manage to deliver an affordable net-zero emissions system. We find that levelised cost of energy decreases due to heat savings, but not for global temperature increases. In both cases, heat pumps become less favourable as surplus electricity is more abundant for heating. Demand-side management through buildings' thermal inertia could shape the heating demand, yet has modest impact on the system configuration. Cost reductions of heat pumps impact resistive heaters substantially, but not the opposite. Cheaper power-to-gas could lower the need for thermal energy storage.
\end{abstract}

\end{frontmatter}

\begin{table}[!t]
\begin{framed}
\nomenclature{GHG}{greenhouse gas}
\nomenclature{PV}{photovoltaics}
\nomenclature{VRES}{variable renewable energy sources}
\nomenclature{LCOE}{levelised cost of energy}
\nomenclature{HVDC}{high voltage direct current}
\nomenclature{PyPSA}{python for power system analysis}
\nomenclature{DSM}{demand-side management}
\nomenclature{HCD}{heating and cooling demand}
\nomenclature{PtH}{power-to-heat}
\nomenclature{PtG}{power-to-gas}
\nomenclature{DAC}{direct air capture}
\nomenclature{CDH}{cooling degree hour}
\nomenclature{HDH}{heating degree hour}
\nomenclature{COP}{coefficient of performance}
\nomenclature{CHP}{combined heat and power}
\nomenclature{CF}{capacity factor}
\nomenclature{PHS}{pumped hydro storage}
\printnomenclature
\end{framed}
\end{table}

\section{Introduction} \label{sec:introduction}
The Special Report on the impacts of global warming of 1.5~\si{\celsius} published by the Intergovernmental Panel on Climate Change (IPCC) states that limiting the global temperature increase to 1.5~\si{\celsius} is possible \cite{SR15}. Yet it requires deep and rapid decarbonisation in all sectors. The global anthropogenic greenhouse gas (GHG) emissions are required to undergo an unprecedented reduction in this century, reaching net zero by 2050. The European Commission released its long-term strategy towards a \co{}-neutral Europe by 2050 \cite{cleanplanet}, in line with the Paris Agreement commitment to limit the temperature increase well below 2~\si{\celsius} and pursue efforts towards 1.5~\si{\celsius}. Available mature low-carbon energy technologies, in particular wind and solar photovoltaic (PV) power plants, are capable of supplying electricity at a large scale. Other sectors apart from electricity, such as transportation, industry, heating and cooling still lack a clear decarbonisation strategy. The far-reaching emissions reductions will benefit from a fully sector-coupled energy system in order to exploit the synergies among sectors \cite{lund2017smart,brown2018synergies,henning2014comprehensive,palzer2014comprehensive}. In this paper, we investigate how changes in climatic, technical and economic assumptions affect the design of a future fossil-free energy system in Europe. The large seasonal variation in heating demand is one of the challenges for the decarbonisation of this sector \cite{zhu2019impact}. However, future heating demand could be significantly impacted by climate change, savings due to building retrofitting, and demand-side management strategies.

Even if we stop emitting anthropogenic GHG today, temperature increase is inevitable, which could impact weather-dependent future energy systems in two ways, supply and demand. Kozarcanin \textit{et al.} \cite{kozarcanin201921st} use weather-driven modelling to investigate the impact of climate change on highly renewable European electricity systems for three distinctive scenarios. They find that climate change could modify the need for dispatchable electricity up to 20\%, but barely affects the benefits of transmission and storage, whose change is below 5\%. Schlott \textit{et al.} \cite{schlott2018impact} explore the impact of climate change on the European electricity systems by techno-economical optimisation assuming no emissions reductions. They find that climate change is expected to increase the correlation length for wind generation. Therefore, PV becomes more cost-competitive and the need for dispatchable energy rises as balancing by reinforcing interconnection among countries is less efficient. Hdidouan and Staffell \cite{hdidouan2017impact} assess future wind capacity factors in Great Britain taking climate change into account. It is concluded that climate change would result in capacity factors increase in some regions but decrease in others. Furthermore, potential temperature increase would very likely decrease overall heat demand while raising cooling demand \cite{EEA19}. Kozarcanin \textit{et al.} \cite{kozarcanin2019impact} find that heating demand decreases by up to 42\% in the most extreme global warming scenarios. They determine the cost-optimal mix of heating technologies and find that heat pumps become more cost-competitive for all the temperature-increase scenarios. Staffell and Pfenninger \cite{staffell2018increasing} show that electricity supply and demand are becoming increasingly weather-dependent, due to higher penetrations of renewable and the rise of heat electrification.

Heating and cooling demand (HCD) is expected to undergo a substantial change due to heat savings. Lund \textit{et al.} \cite{lund2014heat} explore the relation between heat savings and supply in Denmark, and find that the cost-optimal strategy includes 35\% to 53\% savings due to building retrofitting compared to the current level. Likewise, Hansen \cite{hansen2016heat} \textit{et al.} extend the analysis to other European countries and find similar numbers. \cite{agora2018} predicts building retrofitting will decrease final energy demand of buildings in Germany by 44\% in 2050 compared to 2011. Despite its huge potential to facilitate energy system transition, the lack of necessary legislation and control hinder the deployment of heat saving \cite{tommerup2006energy}.

\begin{figure*}[!t]
\centering
\begin{adjustbox}{scale=1,trim=0 8cm 0 0}
\begin{circuitikz}
\draw (6,14.5) to [short,i^=import] (6,13);
\draw [ultra thick] (-9,13) node[anchor=south west]{\textbf{electricity}} -- (9,13);
\draw (7,13) |- +(0,0.5) to [short,i^=export] +(2,0.5);
\draw (-2,13) -- +(0,0.5) node[sground,rotate=180]{};\draw (0,0.5);
\draw (0,14) node[vsourcesinshape, rotate=90](V2){}(V2.left) -- +(0,-0.6);
\draw (1.75,14.3) node{wind, solar PV};
\draw (1.5,13.9) node{run of river};
\draw (1.8,13.5) node{hydro reservoir};
\node[draw,minimum width=1cm,minimum height=0.6cm,anchor=south west] at (-7,13.5){battery, PHS, hydrogen};
\draw (-5,13) to (-5,13.5);
\draw [ultra thick] (-9,10) node[anchor=south west]{\textbf{rural heat}} -- (-3,10) ;
\draw (-8,10) -- +(0,-0.5) node[sground]{};
\node[draw,minimum width=1cm,minimum height=0.6cm,anchor=south west] at (-8,11){synthetic gas};
\draw (-7,13) to [short, i^=\ ] (-7,11.6);\draw (-8,12.2) node{methanation};
\draw (-6.8,11.6) to [short, i^=\ ] (-6.8,13);\draw (-6.2,12.2) node{OCGT};
\draw (-6.9,11) to [short, i^=gas boiler] (-6.9,10);
\draw (-3.5,11.9) node{heat pump};
\draw (-4.5,13) to [short,i^=resistive heater] (-4.5,10); \draw (-3.5,11) node{(individual)};
\node[draw,minimum width=1cm,minimum height=0.6cm,anchor=south west] at (-6.5,8.7){short-term storage};
\draw (-5,10) to (-5,9.3);
\draw [ultra thick] (3,10)  -- (9,10) node[anchor=south east]{\textbf{urban heat}};
\draw (8,10) -- +(0,-0.5) node[sground]{};
\node[draw,minimum width=1cm,minimum height=0.6cm,anchor=south west] at (3.5,11){synthetic gas};
\draw (4.5,13) to [short, i^=\ ] (4.5,11.6);\draw (3.5,12.2) node{methanation};
\draw (4.7,11.6) to [short, i^=\ ] (4.7,13);\draw (5.3,12.4) node{OCGT};\draw (5.2,12) node{CHP};
\draw (4.6,11) to [short, i^=CHP] (4.6,10);\draw (3.7,10.5) node{gas boiler};
\draw (7.5,11.9) node{heat pump};
\draw (6.5,13) to [short,i^=resistive heater] (6.5,10);
\draw (7.3,11) node{(central)};
\node[draw,minimum width=1cm,minimum height=0.6cm,anchor=south west] at (3.5,8.7){long-term storage};
\draw (5,10) to (5,9.3);
\draw [ultra thick] (-2,10) -- (2,10) node[anchor=south east]{\textbf{cooling}};
\draw (0,10) -- +(0,-0.5) node[sground]{};
\draw (-1,13) to [short,i^=heat pump] (-1,10);
\draw (0.2,11) node{(cooling mode)};
\end{circuitikz}
\end{adjustbox}
\caption{Energy flow diagram of one node, representing a European country. Each node is divided into four buses: one electricity bus, two heat buses named rural and urban according to population density, as well as one cooling bus. Loads (triangles), generators (circles), storage units (rectangles), transmission lines and energy converters are attached to buses. Individual heat pumps can operate in two modes, providing both heating and cooling. The synthetic gas is consumed by conventional units, such as OCGT, CHP and gas boilers.} 
\label{fig:flow}
\end{figure*}
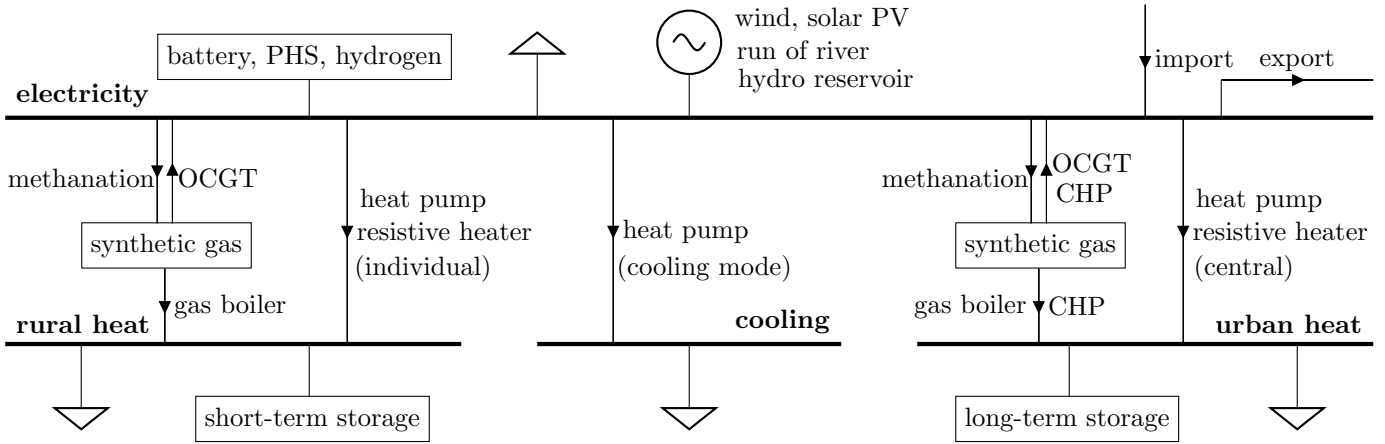

Demand-side management (DSM) could alter the shape of consumers' pattern, in order to decrease the demand at peak times. Among various implementations of DSM, installing additional equipment, \textit{e.g.}, thermal energy storage, turns out to be highly efficient \cite{arteconi2012state,arteconi2013domestic}. Long-term thermal energy storage represented by large hot water tanks in district heating systems shapes the seasonal variation of HCD \cite{brown2018synergies}. Short-term thermal energy storage, as individual hot water tanks, could smooth out the daily variation \cite{VICTORIA2019111977}. However, additional equipment require space and investment, making it less attractive. An alternative solution is to utilise the buildings' structure itself, exploiting the potential of thermal mass without sacrificing thermal comfort. Reynders \textit{et al}. \cite{reynders2013potential} analyse the potential of structure thermal mass for DSM in order to avoid a strong mismatch between electricity production and consumption in residential buildings. A large potential is found in using the thermal mass as short-term storage to shift the peak electricity demand. In \cite{le2016energy}, the flexibility potentials are estimated by increasing or decreasing the set-point temperature in two different types of buildings. Poorly-insulated buildings are able to maintain the thermal comfort zone up to 5 hours after switching off the heating, whereas well-insulated buildings can endure for 24 hours.

Previously, some authors have examined the different aspects of uncertainties. Schlachtberger \textit{et al.} \cite{schlachtberger2018cost} investigate the influence of weather data, cost parameters and policy constraints on a highly renewable European electricity system. A considerable robustness of system costs to weather data and cost assumptions is observed. Victoria \textit{et al.} \cite{VICTORIA2019111977} show that different cost assumptions for storage technologies have a significant impact on the types and capacities of storage deployed, but low impact on the system costs. Collins \textit{et al.} \cite{collins2018impacts} analyse the impact of long-term weather patterns on European electricity system. A 5-fold increase has been revealed in terms of inter-annual variability of \co{} emissions by 2030 compared with 2015.

In \cite{zhu2019impact}, we focus on the role of \co{} prices for a highly decarbonised coupled electricity and heating system in Europe. We find out that not only a renewable target is necessary, but also a \co{} tax is required to incentivise the cost-optimal system configuration. For the cost optimal configuration with 95\% \co{} reductions relative to 1990, most investments go into variable renewable energy sources (VRES) and power-to-heat (PtH) installations, and heating sector is supplied mostly by heat pumps. Compared to electricity demand, HCD has a larger seasonal variation, which would decrease the benefits of high-efficiency but high capital cost technologies, such as heat pumps. HCD could be influenced by three causes: temperature increase because of climate change, heat saving from building retrofitting and demand-side management through buildings' thermal inertia. In this study, we evaluate the impact of climatic, technical and economic uncertainties on the coupled electricity, heating and cooling European energy system, to address the following research questions:

\begin{itemize}
\item How does a different HCD alter the optimal system configuration, under the circumstances of temperature increases, heat savings or demand-side management strategies?
\item What would be the impact of cost reductions on the key components?
\end{itemize}

The paper is organised as follows. Section \ref{sec:methods} concisely describes the model and definitions of different scenarios under analysis. A detailed model description can be found in \ref{sec:appendix}. Section \ref{sec:results} presents the results of this study and the subsequent Section \ref{sec:discussions} discusses the main findings as well as the limitations of this analysis. Finally, Section \ref{sec:conclusions} draws the main conclusions.

\section{Methods} \label{sec:methods}

\begin{table*}[!b]
\begin{threeparttable}
\caption{Cost, efficiency and lifetime assumptions for the key technologies} \label{tab:cost parameters}
\centering
\begin{tabularx}{\textwidth}{l|c|c|c|c|c|c}
\toprule
Technology&Overnight cost[\euro]&Unit&FOM\tnote{a}[\%/a]&Lifetime[a]&CF/Efficiency\tnote{b}&source\\
\midrule
Onshore wind\tnote{c}&910&kW\el&3.3&30&0.23[0.07-0.33]&\cite{DEA2019}\\
Offshore wind\tnote{c}&2506&kW\el&3&25&0.31[0.09-0.51]&\cite{schroder2013current}\\
Solar PV\tnote{c}&575&kW\el&2.5&25&0.13[0.06-0.19]&\cite{vartiainen2017true}\\
OCGT\tnote{d}&560&kW\el&3.3&25&0.39&\cite{DEA2019}\\
CHP\tnote{d}&600&kW\th&3.0&25&0.47&\cite{henning2014comprehensive}\\
Gas boiler\tnote{d,e}&175/63&kW\th&1.5&20&0.9&\cite{palzer2016sektorubergreifende}\\
Resistive heater&100&kW\th&2&20&0.9&\cite{schaber2014integration}\\
Heat pump\tnote{e}&1400/933&kW\th&3.5&20&[3.03-3.79]/[2.73-3.04]&\cite{palzer2016sektorubergreifende}\\
Battery inverter&310&kW\el&3&20&0.81&\cite{budischak2013cost}\\
Battery storage&144.6&kWh&0&15&-&\cite{budischak2013cost}\\
Electrolysis&350&kW\el&4&18&0.8&\cite{palzer2016sektorubergreifende}\\
Fuel cell&339&kW\el&3&20&0.58&\cite{budischak2013cost}\\
Hydrogen storage&8.4&kWh&0&20&-&\cite{budischak2013cost}\\
Methanation+DAC&1000&kW\h2&3&25&0.6&\cite{palzer2016sektorubergreifende}\\
Hot water tank\tnote{f}&860/30&$m^3$&1&20/40&3/180~days&\cite{henning2014comprehensive}\\
HVDC lines&400&MWkm&2&40&1&\cite{hagspiel2014cost}\\
\bottomrule
\end{tabularx}
\begin{tablenotes}
\footnotesize
\item [a] Fixed Operation and Maintenance (FOM) costs are given as a percentage of the overnight cost per year.
\item [b] Capacity Factor (CF) only applies to renewables, while efficiency only to generators and converters.
\item [c] Capacity Factor varies in different countries due to weather condition. The number in front indicates the average of CF weighted by demand, while the numbers in brackets show the range of CF for different countries. Solar PV is split 50-50\% between rooftop and utility-scaled power plants. The impacts of this assumption is limited as discussed in \cite{victoria2019pv}.
\item [d] OCGT, CHP and gas boilers consume synthetic gas generated from methanation.
\item [e] Gas boilers and heat pumps have different costs and efficiencies for individual (numbers in front) and centralised (numbers behind) systems. The efficiency of heat pumps, also known as the coefficient of performance, varies with temperature.
\item [f] \label{tab: note f} Heat losses of central hot water tanks are much lower than individual, 1/4320 (180 days) and 1/72 (3 days) per hour respectively.
\end{tablenotes}
\end{threeparttable}
\end{table*}

\subsection{Model}
In this paper, we model the fossil-free European energy system, where electricity, heating and cooling sectors are coupled. The model is implemented as a linear techno-economical optimisation assuming perfect foresight and long-term market equilibrium, which ensures that the costs of optimised technologies are exactly recovered by market revenues. We use the open-source framework python for power system analysis (PyPSA) \cite{brown2017pypsa} and the PyPSA-Eur-Sec-30 instance introduced in \cite{brown2018synergies}.

The objective function is expressed as the total annualised system costs including capital and marginal costs (Equation \ref{eq:objective}). Technical and physical constraints such as hourly supply of inelastic demand and \co{} emissions, are imposed (Equations \ref{eq:energybalance} - \ref{eq:alpha}). In each country, the sum of wind and solar PV generation is proportional to its annual demand, but the wind/solar PV mix is optimised according to diverse renewable resource (Equation \ref{eq:alpha}). It depicts a plausible future scenario where the European countries need to be relatively self-sufficient in terms of renewable supply. The same constraint is imposed in \cite{zhu2019impact}, but not in \cite{brown2018synergies}. A fossil-free system allows consuming synthetic gas only after it has been produced, via power-to-gas where direct air captured \co{} (DAC) and electrolysed hydrogen are combined to produce methane. The Lagrange/Karush-Kuhn-Tucker multiplier $\mu_{CO_2}$ corresponding to that constraint (Equation \ref{eq:CO2 price}) indicates the necessary \co{} tax to obtain net-zero emissions in an open market. Since a fossil-free energy system is considered to be in the far future, the currently existing generators are not included in this model, \textit{i.e.}, we use greenfield optimisation.

\begin{figure}[!t]
\centering
\includegraphics[trim=0 0cm 0 0,width=\picsize\linewidth,clip=true]{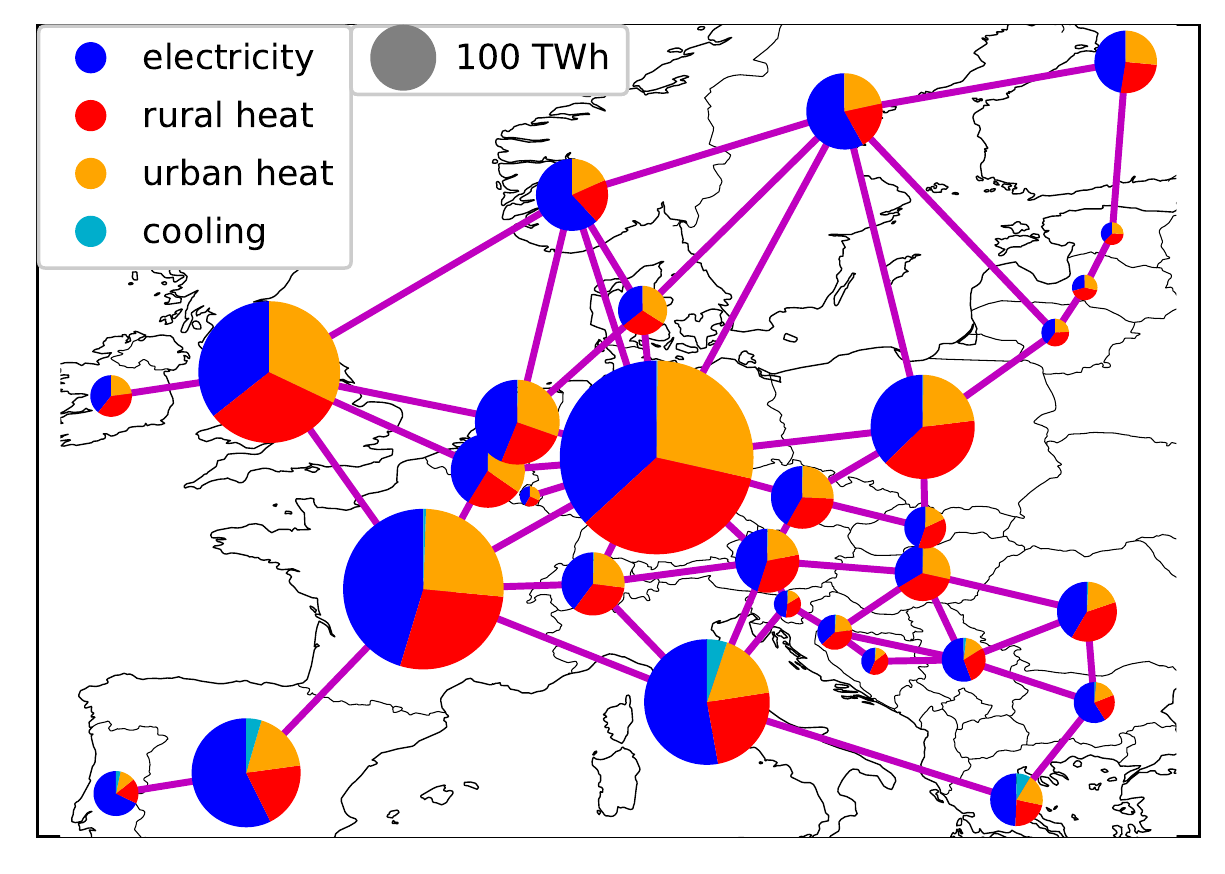}
\caption{Annual national demand and transmission grid map. Demand consists of electricity (2851~TWh\el), heating in high population-density areas (urban heat, 1624~TWh\th), heating in low population-density areas (rural heat, 1939~TWh\th) and cooling (70~TWh\th). Transmission lines include existing and under-construction lines.}
\label{fig:topology and demand}
\end{figure}

Each of the 30 European countries, \textit{i.e.}, 28 European Union member states excluding Malta and Cyprus but including Norway, Switzerland, Serbia, and Bosnia-Herzegovina, is represented by a single node. Neighbouring countries are connected through cross-border High Voltage Direct Current (HVDC) lines, see Figure \ref{fig:topology and demand}, where transmission capacities are optimised at given investment costs. Within each country, electricity, heating and cooling sectors are coupled, as shown in Figure \ref{fig:flow}. 

The electricity demand is taken from historical values of 2015 provided by ENTSO-e \cite{OPSD}. The load can be supplied by renewable generators, \textit{i.e.}, wind, solar PV and hydro electricity. Synthetic gas could be used in OCGT and CHP to supply residual load. Electric storage in the forms of static batteries, pumped hydro storage and hydrogen storage in overground steel tanks can store electricity for later usage. The cost assumptions for key technologies are shown in Table \ref{tab:cost parameters}. In order to acknowledge potential cost decreasing, particularly for wind and solar PV, while avoiding uncertainties of long-term projection, the cost assumptions are taken from the predictions for 2030, assuming a discount rate of 7\% for the annualised overnight cost.

\subsubsection*{Heating}
Turning to the heating sector, only residential and commercial sectors are considered, which can be further divided into space heating and hot water demand. The profiles of space heating are approximated by heating degree hour (HDH), assuming the demand rises linearly according to ambient temperature $T^{amb}$ below a threshold $T^{th}$ in country $i$,
\begin{equation}
HDH_{i,t} = (T_{i,t}^{th} - T_{i,t}^{amb})^+  \label{eq:HDH}
\end{equation}
where the sign `+' indicates that only positive values are counted, and the threshold temperature is assumed to be 17~\si{\celsius}. The time series is then scaled based on the annual demand for domestic space heating in 2015 \cite{HRE}. A country-specific constant hourly value for the hot water consumption, obtained from the same database, is added to compute the total heating demand time series representative for every country. Heating demand can be supplied by power-to-heat technologies, \textit{i.e.}, heat pumps and resistive heaters, and dispatchable backup, \textit{i.e.}, gas boilers and CHP powered by synthetic gas. In order to capture the low performance in winter, the coefficients of performance (COP) of heat pumps vary with ambient temperature \cite{staffell2012review}. The synthetic gas is produced through the Sabatier process from electrolysed hydrogen and direct air captured \co{}. The alternative source of \co{} could be other cheaper carbon sources, such as the industry sector, but this option is not considered in this paper. Power-to-gas (PtG), in this paper, refers to both processes using electricity to produce gas, whether hydrogen or synthetic gas. In high population density areas, central thermal energy storage, \textit{i.e.}, large hot water tanks, can be installed and connected to district heating systems. In low population density areas, individual thermal energy storage is deployed, see \cite{brown2018synergies}. 

\subsubsection*{Cooling}
Instead of aggregating the cooling demand into electricity, it is treated as a separate bus in order to capture its potential rise due to temperature increase. The cooling demand only consists of space cooling, and the profiles are modelled through cooling degree hour (CDH), assuming the demand rises linearly according to ambient temperature $T^{amb}$ above a threshold $T^{th}$
\begin{equation}
CDH_i = \sum_{t}(T_{i,t}^{amb} - T_{i,t}^{th})^+, \hspace{5mm} \forall i \label{eq:CDH}
\end{equation}
where the threshold temperature is assumed to be 24~\si{\celsius}. Likewise, the time series is scaled according to annual cooling demand in 2015. Cooling can only be provided by reversing heat pumps into cooling mode \cite{staffell2012review}, assuming a fixed COP of 3.

\subsection{Scenarios}
The label `Base' refers to the baseline scenario without incorporating the climatic, technical or economic changes as described below. 

\begin{figure}[!t]
	\centering
	\includegraphics[trim=0 0cm 0 0,width=\picsize\linewidth,clip=true]{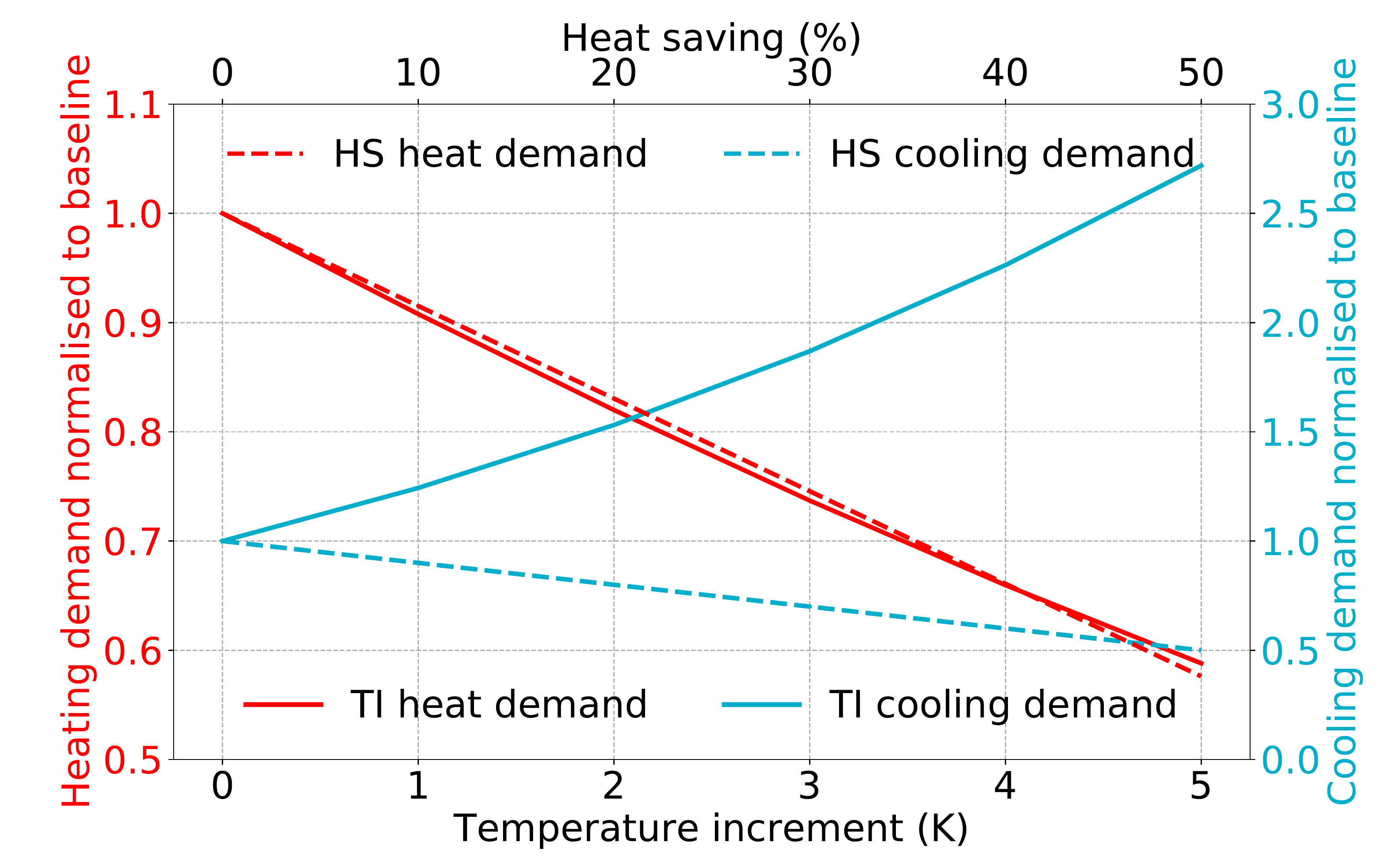}
	\caption{Normalised to baseline Europe-aggregated heating and cooling demand as a function of different levels of temperature increase (labelled by `TI') and heat saving (labelled by `HS'). Temperature increase of 0~\si{\kelvin} and heat saving of 0\% are equivalent to baseline.}
	\label{fig:HCD TI HS}
\end{figure}

\subsubsection*{Climatic uncertainties}
To investigate the impact of climate change on the energy system, a simplified approach is introduced. The ambient temperature increments are assumed to be homogeneous both in space and time, \textit{i.e.}, certain degrees $\Delta T$ are added to all the countries for all the hours. The resulting ambient temperature would be,
\begin{equation*}
T_{i,t}^{amb}+\Delta T, \hspace{5mm} \forall i, t
\end{equation*}
where $\Delta T$ ranges from 1 to 5~\si{\kelvin}. Since space heating and cooling demands are modelled by HDH and CDH approximations (Equations \ref{eq:HDH} - \ref{eq:CDH}), temperature increments would lead to lower heating demand and higher cooling demand. The solid lines in Figure \ref{fig:HCD TI HS} provide an overview of changes in annual heating and cooling demand versus temperature increments. It is also worth mentioning that as temperature increases, heat pumps have higher efficiencies. For instance, the averaged COP for baseline is 3.39, but the number becomes 3.81 for temperature increase of 5~\si{\kelvin}.

\subsubsection*{Technical uncertainties}
The technical aspect is investigated in two scenarios: heat savings (HS) and demand-side management (DSM).

To model the heat savings due to potential improve of energy efficiency in buildings, homogeneous reduction levels $\Delta H$ are applied on space heating and cooling demand from baseline spatially as well as temporally. The new heating demand time series for all the countries become,
\begin{equation*}
(100\% - \Delta H)\cdot d_{n,t}, \hspace{3mm} n\in \text{space heating\&cooling, } \forall t
\end{equation*}
We consider reduction levels from 10 up to 50\%. It must be noted that the expenses of heat savings are not included in the optimisation. The dashed lines in Figure \ref{fig:HCD TI HS} present the effects of heat savings on heating and cooling demand. 

\begin{figure}[!b]
	\centering
	\includegraphics[trim=0 0cm 0 0,width=\picsize\linewidth,clip=true]{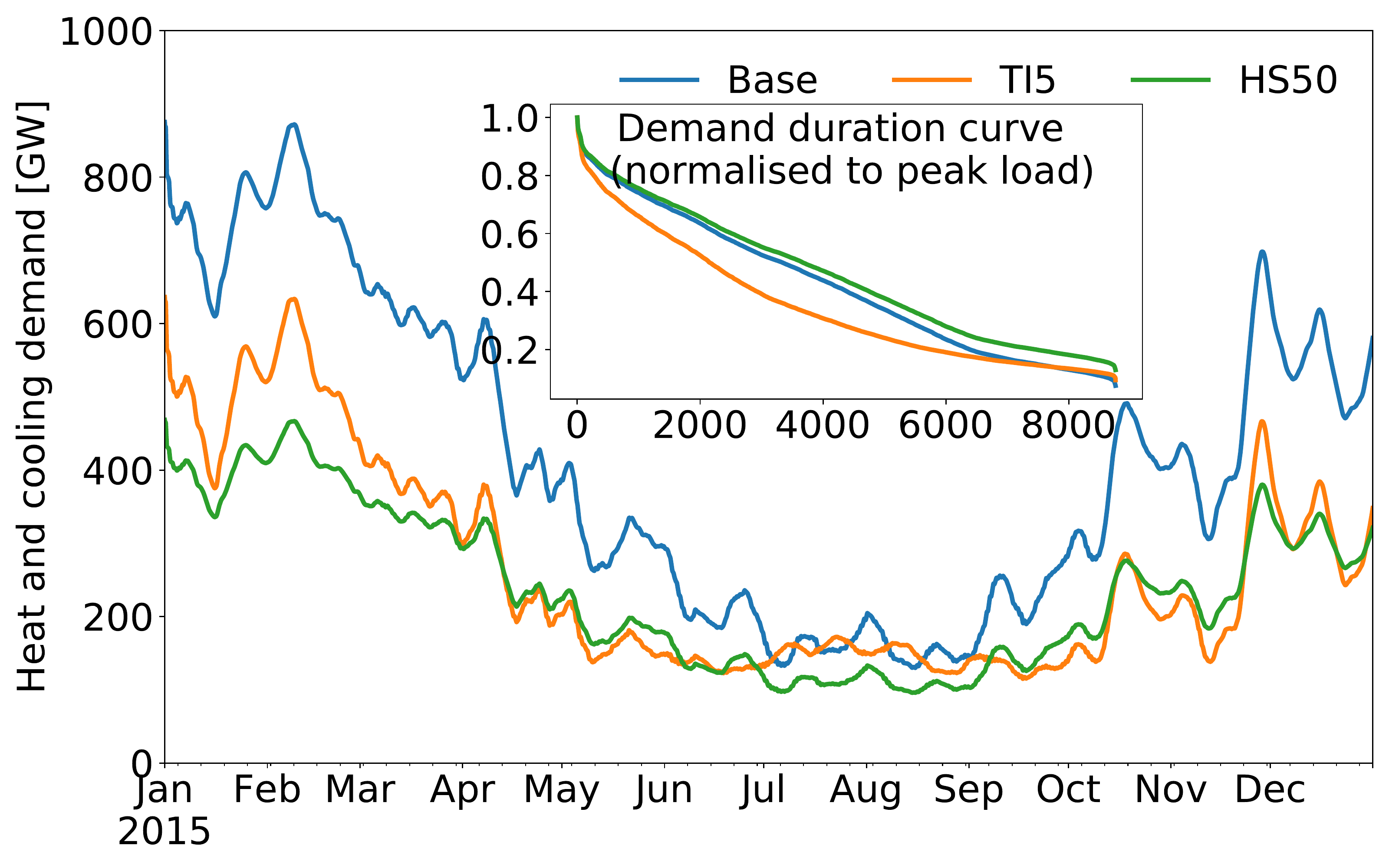}
	\caption{Europe-aggregated heating and cooling demand comparison between baseline (labelled by `Base'), temperature increase of 5~\si{\kelvin} (labelled by `TI5') and heat saving of 50\% (labelled by `HS50'). The hourly loads have been averaged to one week. The inset shows the demand duration curves normalised to the corresponding peak load.}
	\label{fig:HCD profile}
\end{figure}

Temperature increments and heat savings do not only change the total HCD, but also modify the seasonal variation to some extent. Figure \ref{fig:HCD profile} shows the one-week-smoothed hourly profiles of HCD for baseline, temperature increase of 5~\si{\kelvin} and heat saving of 50\%. The impact of temperature increases on the HCD are mixed: the heating demand decreases in winter while cooling demand increases during summer. Temperature increases differentiate the duration curve substantially from the baseline, resulting in a flatter tail, see the inset of Figure \ref{fig:HCD profile}. Comparatively, heat savings modify the space heating and cooling demand uniformly, hence preserving the profile of HCD for most of the time.

Among various methods for demand-side management for the heating sector, this paper introduces a simple way by utilising buildings' thermal mass as short-term storage without sacrificing the indoor comfort. A zero-cost, capacity-fixed thermal energy storage is attached to each heat bus, which allows for charging extra heating during off-peak hours and discharging for the peak. The power capacity $G_{n,s}$ is assumed to be the average heating demand in each heat bus, and the energy capacity $E_{n,s}$ is obtained by multiplying power capacity with certain time constant $\tau$, 
\begin{equation*}
E_{n,s} = \tau \cdot G_{n,s}, \hspace{5mm} s \in \text{DSM storage}
\end{equation*}
where $n$ refers to either urban or rural heat bus. Time constants up to 10 hours are considered in this paper, in order to sustain the indoor thermal comfort. The extra heating stored in the structure thermal mass has high heat losses, \textit{i.e.}, $1/\tau$ per hour. Comparatively, individual and central thermal energy storage's heat losses are significantly lower, see the note f in Table \ref{tab:cost parameters}.

\subsubsection*{Economic uncertainties}
The economic aspect is analysed in terms of three key components in supplying heating and cooling demand, \textit{i.e.}, heat pumps (HP), resistive heaters (RH) and power-to-gas (PtG). 

Heat pumps provide most of the thermal energy among all the heating suppliers in the baseline scenario, see Figure \ref{fig:energy flow Base}. Therefore, the annualised capital costs of heat pumps play a vital role in the highly decarbonised electricity, heating and cooling coupled system. Certain reduction rates $\Delta C_{HP}$ are assumed for heat pump costs, resulting in new annualised capital costs as,
\begin{equation*}
(100\%-\Delta C_{HP})\cdot c_{n,s}, \hspace{5mm} s \in \text{heat pumps}
\end{equation*}
Reduction rates from 10 to 50\% are taken into account in this study.

Comparatively, resistive heaters are much cheaper to install but have lower efficiencies. In a similar manner, the cost reductions, $\Delta C_{RH}$, are applied to annualised capital costs of resistive heaters, 
\begin{equation*}
(100\%-\Delta C_{RH})\cdot c_{n,s}, \hspace{5mm} s \in \text{resistive heaters}
\end{equation*}
where reduction rates range from 10 to 50\%.

Apart from power-to-heat technologies, dispatchable energy using synthetic gas plays an important role in supplying heating during peak hours. Cost reduction levels are applied to capital costs linked to the power-to-gas, \textit{i.e.}, electrolysis and methanation,
\begin{equation*}
(100\%-\Delta C_{PtG})\cdot c_{n,s}, \hspace{5mm} s \in \text{\{electrolysis, methanation\}}
\end{equation*}
Likewise, reduction rates from 10 up to 50\% are considered in this study.

\subsection{Key metrics}
The impact of climatic, technical and economic uncertainties is evaluated in terms of three sets of key metrics. The first set of key metrics consists of system cost and LCOE, which measure the affordability of energy system. The system cost is the objective function of techno-economical optimisation (Equation \ref{eq:objective}). LCOE is calculated as system cost divided by the total energy demand. 

The second set of key metrics looks at the system configuration, \textit{i.e.}, technology composition, wind/solar PV mix (Equation \ref{eq:alpha EU}), VRES penetration (Equation \ref{eq:gamma}) and thermal penetration (Equation \ref{eq:thermal penetration}). Thermal penetrations are defined similarly to VRES penetrations, but in terms of heating and cooling provided by a certain thermal technology. The thermal penetration of technology $s$ is calculated as,
\begin{equation}
\frac{\text{Thermal energy}_s}{\text{Thermal energy}_\text{in total}}, \hspace{3 mm} s\in\text{thermal technologies} \label{eq:thermal penetration}
\end{equation}

The last set of key metrics investigates the operation of energy system, such as heating and cooling supply time series, VRES curtailment, utilisation factor. Those provide a more detailed and practical overview on how the different system components collaborate.

\section{Results}\label{sec:results}

\begin{figure}[!b]
	\centering
	\includegraphics[trim=0 0cm 0 0,width=\linewidth,clip=true]{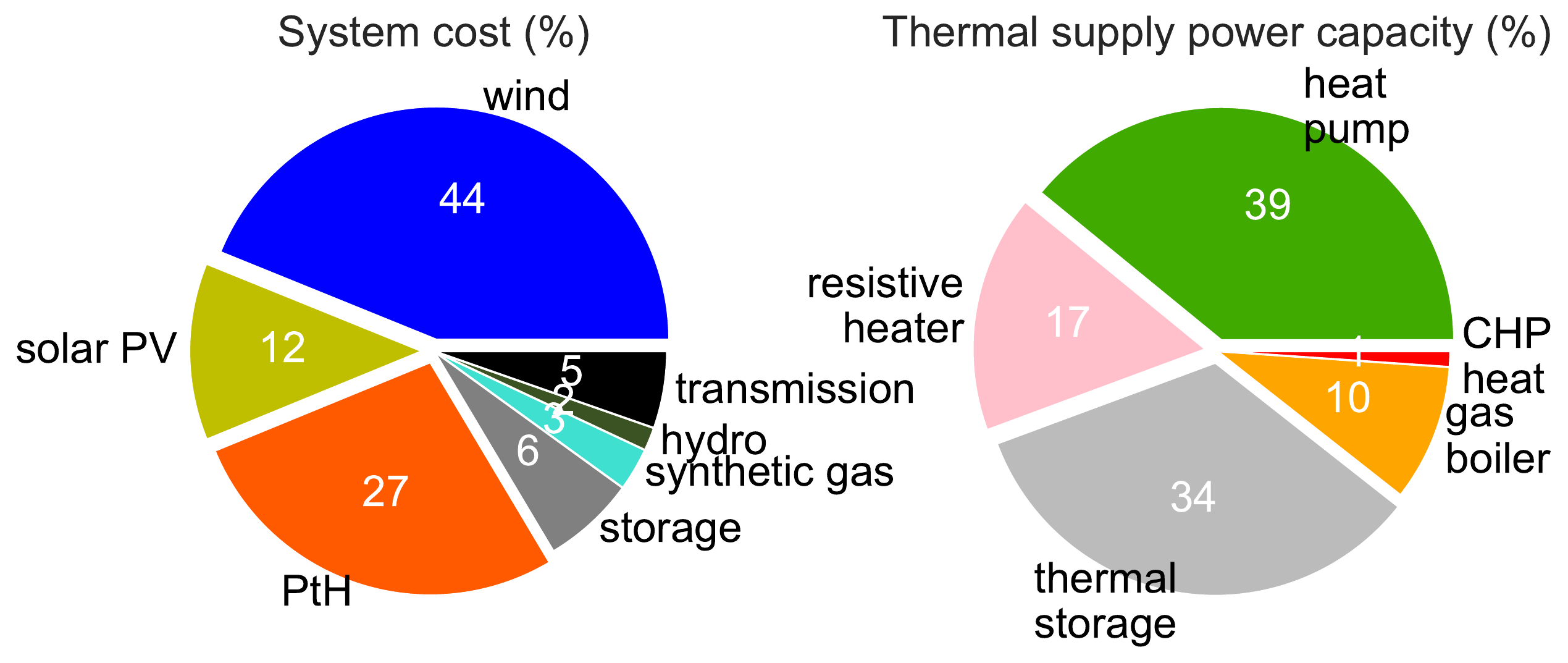}
	\caption{System cost (left) and thermal supply power capacity (right) compositions for baseline. White numbers indicate percentage shares for different technologies.}
	\label{fig:base}
\end{figure}

\subsection{Baseline: net-zero emissions system}

\begin{figure*}[!t]
	\centering
	\includegraphics[trim=0 0 0 0,width=\linewidth,clip=true]{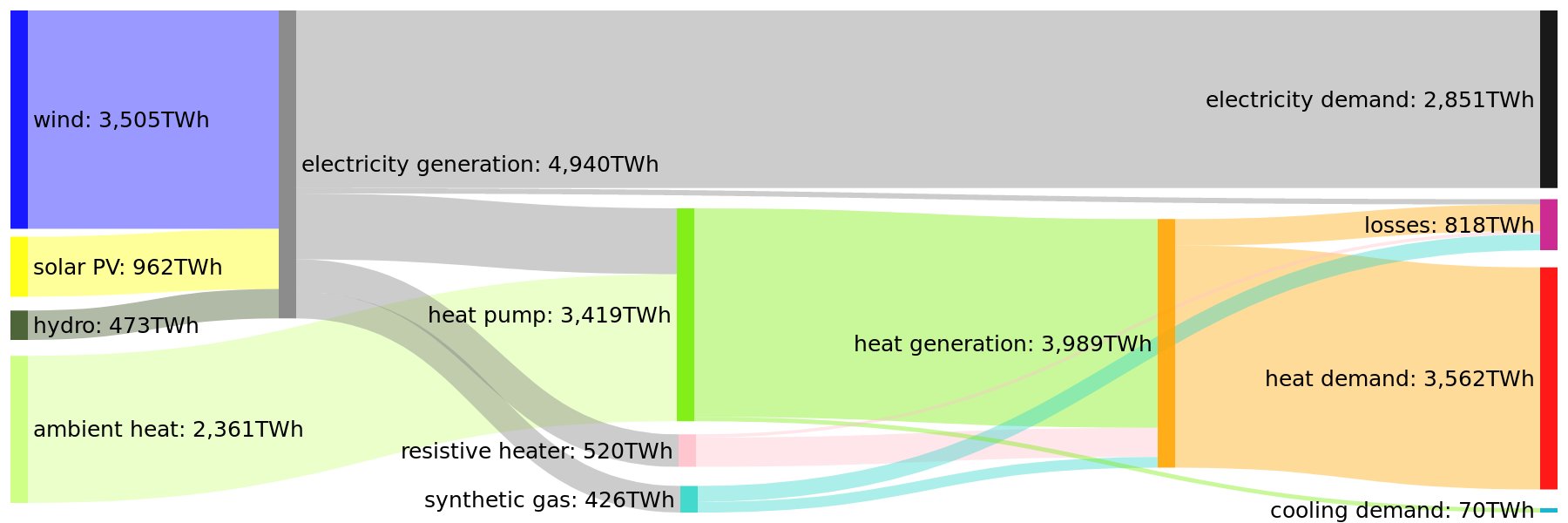}
	\caption{Energy flow for baseline. Synthetic gas only includes the part where power is converted to heating. Losses include conversion and storage losses, but exclude transmission losses.}
	\label{fig:energy flow Base}
\end{figure*}
The coupled electricity, heating and cooling system costs 422~billion~\EUR annually. The cost shares of different technologies are shown in the left plot of Figure \ref{fig:base}. Being the main energy suppliers of the system, the sum of wind and solar PV accounts for more than half of the system cost. Wind turbines generate 71\% (3505 out of 4940~TWh\el) and solar PV 19\% (962 out of 4940~TWh\el) of electricity annually, see Figure \ref{fig:energy flow Base}. The annual VRES curtailment is lower than 1\% of total VRES production. The remaining 10\% of electricity is covered by hydroelectricity from hydro reservoir and run of river . Roughly a quarter of total expenditure goes to power-to-heat technologies, \textit{i.e.}, heat pumps and resistive heaters. The rest of the cost is related to balancing technologies: cross-border transmission to smooth spatial variations and storage for temporal smoothing.

\begin{figure}[!b]
	\centering
	\includegraphics[trim=0 0cm 0 0,width=\picsize\linewidth,clip=true]{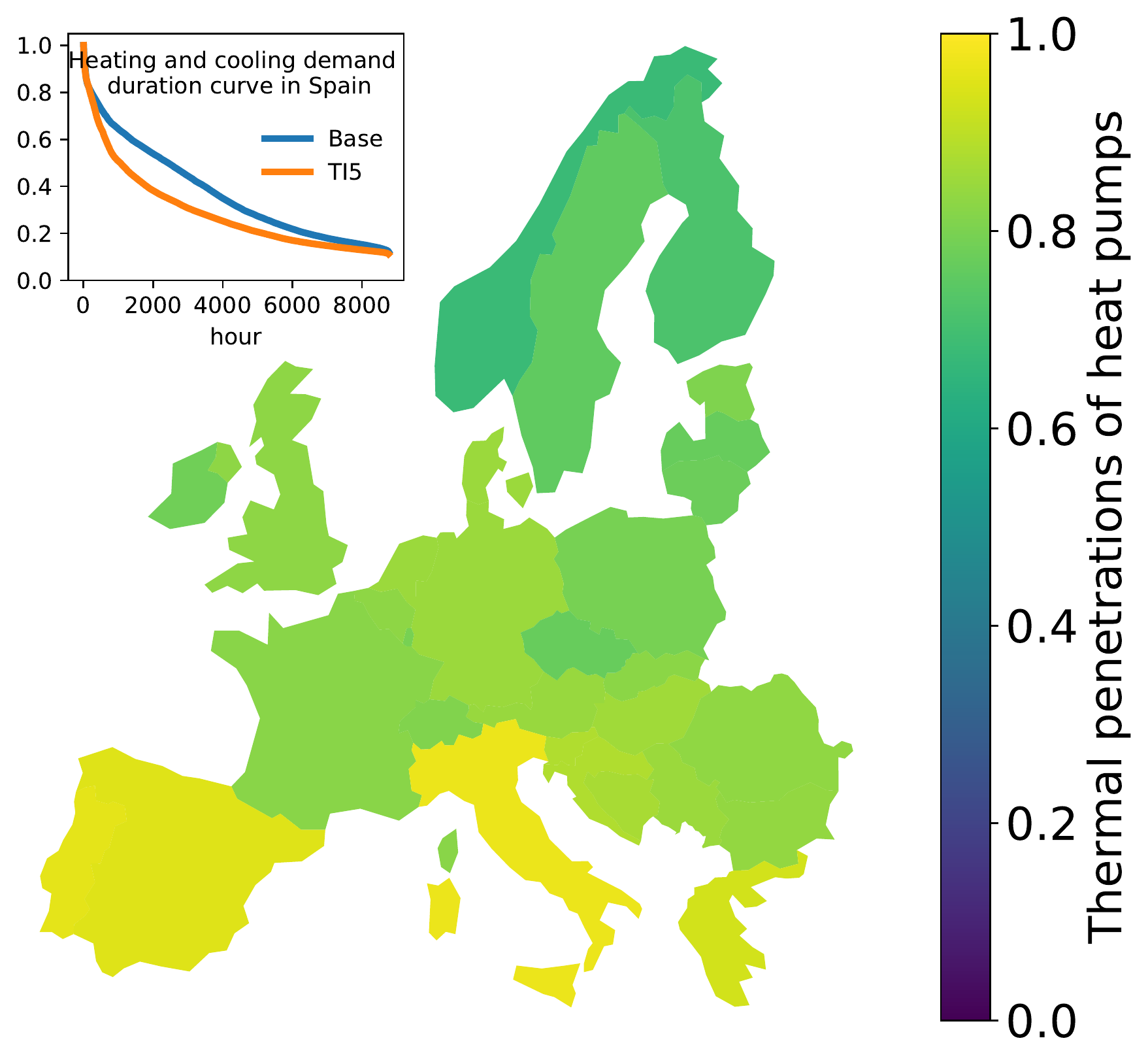}
	\caption{Thermal penetrations of heat pumps for baseline. The inset shows the normalised heating and cooling demand duration curve for Spain comparing temperature increase by 5~\si{\kelvin} (labelled by `TI5') to baseline (labelled by `Base').}
	\label{fig:spatial HP Base}
\end{figure}

The thermal supply capacities are shown in the right plot of Figure \ref{fig:base}. Despite being the most expensive heating technology, heat pumps account for approximately 40\% of total thermal supply capacities (2322~GW\th), and convert 1059~TWh\el{} electricity into 3349 and 70~TWh\th{} of heating and cooling respectively, see Figure \ref{fig:energy flow Base}. Thanks to its high-efficiency, with an averaged COP higher than 3, heat pumps are the most cost-effective thermal technologies, and they supply the majority of heating and cooling for all the European countries, see Figure \ref{fig:spatial HP Base}. Southern countries have higher thermal penetrations of heat pumps, mainly due to the fact that seasonal variations of HCD in northern countries are higher. For instance, optimal thermal penetration of heat pumps is 0.95 for Spain, while 0.68 for Norway. The HCD duration curve of Spain, the blue curve in the inset of Figure \ref{fig:spatial HP Base}, shows a flatter tail compared to the duration curve for Norway depicted in the inset of Figure \ref{fig:spatial HP TI5}. 

Higher seasonal variations of HCD require other thermal technologies, such as resistive heaters, which have lower efficiency but cost significantly less to install. Compared to heat pumps, resistive heaters hold roughly half of thermal supply capacities, but only provide less than 15\% of heating demand. Although \co{} emissions are kept net-zero in the model, gas-to-heat technologies, \textit{i.e.}, gas boilers and CHP heat, still account for about 10\% of the thermal supply capacities and provide 172~TWh\th{} of heating. This result highlights the need of dispatchable backup for heating to secure supply at every hour. Accounting for one third of thermal supply capacities, thermal energy storage in the form of hot water tanks is able to smooth out temporal variations not only for a few days, but also seasonally.

\begin{figure}[!b]
	\centering
	\includegraphics[trim=0 0cm 0 0,width=\picsize\linewidth,clip=true]{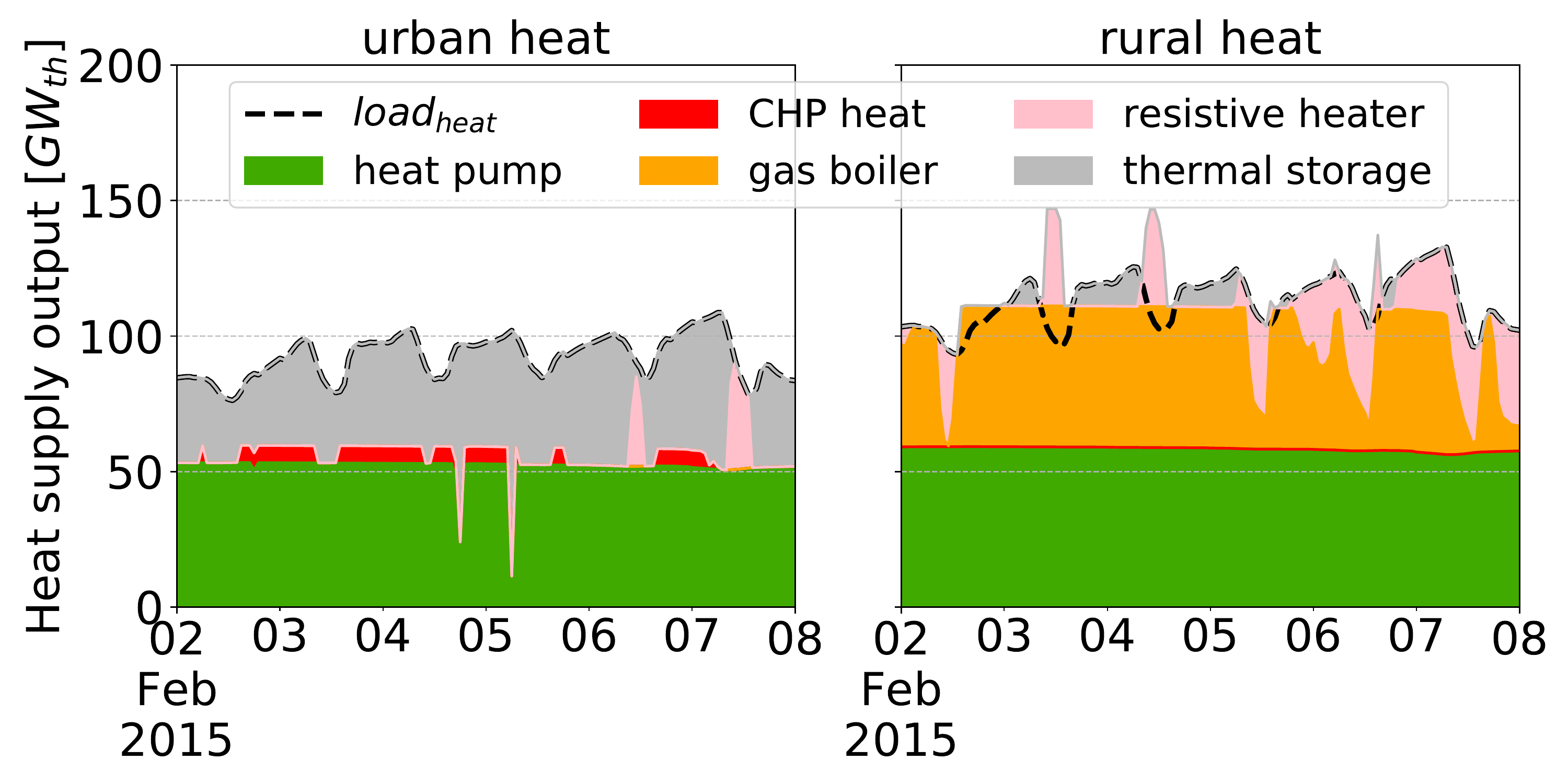}
	\caption{Heating supply output of urban (left) and rural (right) heat buses during a winter week in Germany for baseline.}
	\label{fig:thermal supply base}
\end{figure}

Figure \ref{fig:thermal supply base} shows the heating supply time series during a cold winter week for urban and rural heat buses in Germany. Heat pumps supply most of the heating at a rather constant output for both buses. For urban areas, the residual load is mainly covered from thermal energy storage discharge. By contrast, due to the fact that individual thermal energy storage cannot provide seasonal balancing, rural areas rely heavily on consuming synthetic gas through gas boilers. It is not only for covering the residual load, but also for storing the generated heating for later usage, as seeing a few hours of burning synthetic gas more than the demand needs. The cooperation between synthetic gas and short-term thermal energy storage in rural areas smooths out diurnal variations to some extent, thus lowering the required capacities of power-to-heat technologies. Utilisation factors, calculated by the average operating power divided by nominal capacity, are 0.62, 0.13, 0.07 and 0.06 for heat pumps, resistive heaters, CHP heat and gas boilers respectively, reflecting the different roles that thermal units play in the heating and cooling sector. The synergy between long-term storage provided by synthetic gas and large hot water tanks, inexpensive resistive heaters and high-efficient heat pumps manage to carry out a coupled fossil-free electricity, heating and cooling system.

\subsection{Sensitivity to temperature increases}\label{sec:TI}
\begin{figure}[!b]
	\centering
	\includegraphics[trim=0 0cm 0 0,width=\picsize\linewidth,clip=true]{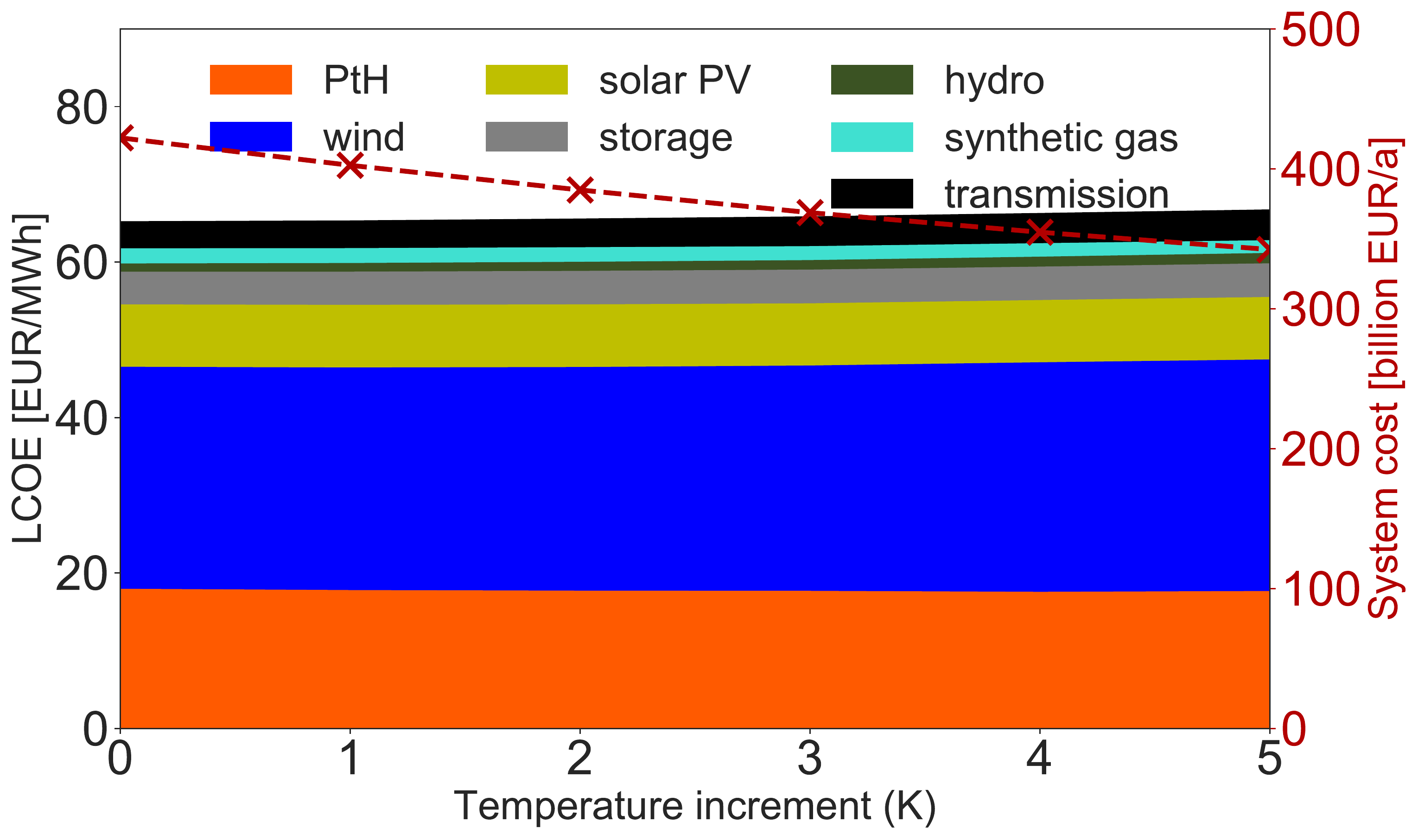}
	\caption{LCOE compositions for temperature increases up to 5\si{\kelvin}. The red dashed line indicates the annualised system costs. Temperature increase of 0~\si{\kelvin} is equivalent to baseline.}
	\label{fig:TI system cost}
\end{figure}

As temperature increases, the compositions and sum of LCOE remain almost the same compared to baseline, see Figure \ref{fig:TI system cost}. In spite of much lower heating demand, the LCOE contribution of power-to-heat does not decrease. The system costs, represented by the red dashed line, drop almost linearly since the total demand falls. Temperature increases do not result in low-priced energy, but reduce the system costs due to diminished demand in heating. 

\begin{figure}[!t]
	\centering
	\includegraphics[trim=0 0cm 0 0,width=\linewidth,clip=true]{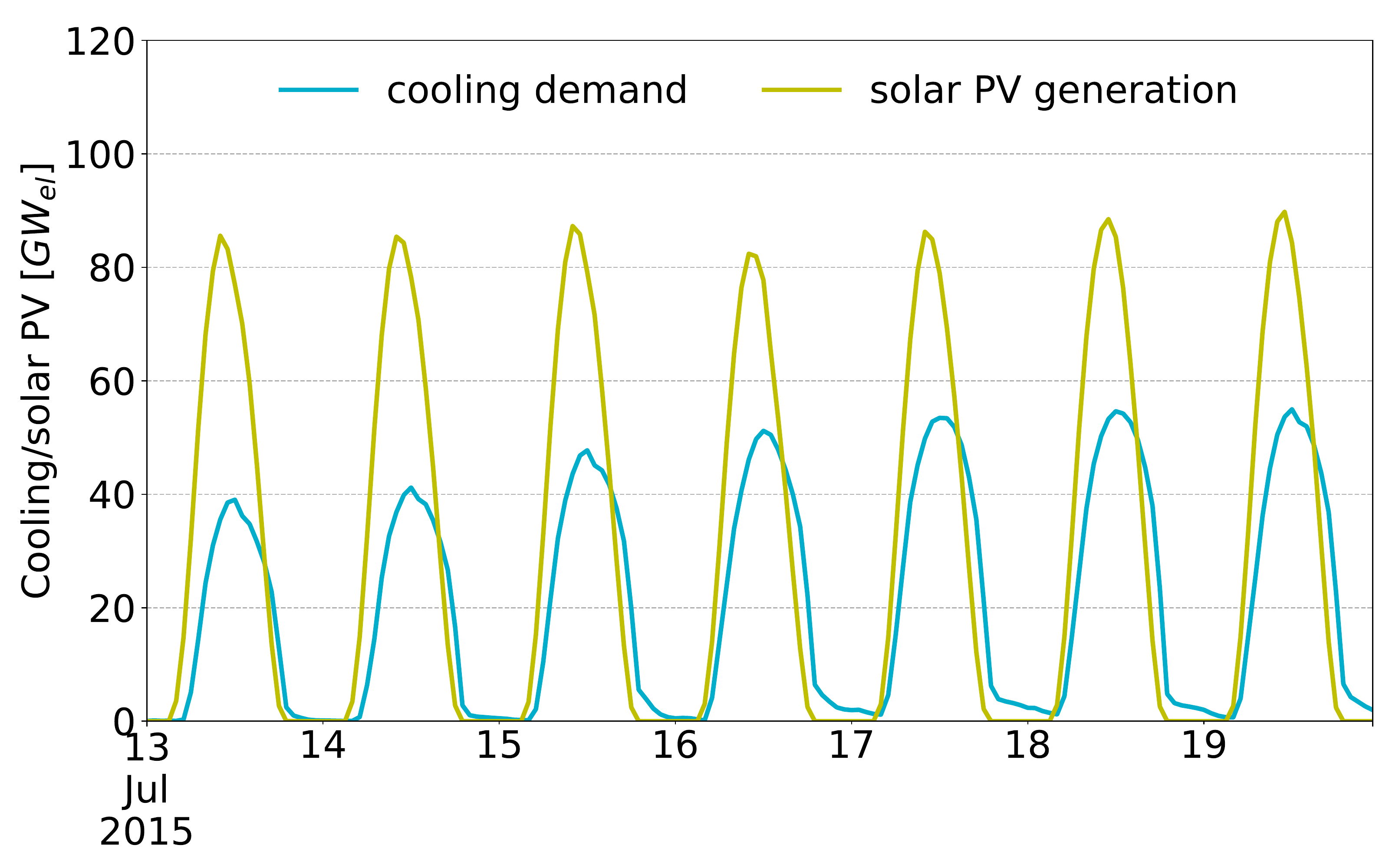}
	\caption{Time series of cooling demand and solar PV generation during a summer week in Italy for the scenario of temperature increase of 5~\si{\kelvin}.}
	\label{fig:cooling and solar}
\end{figure}

The roles of wind and solar PV have been strengthened as the VRES penetrations rise by a small margin. Even though the wind/solar PV mixes at the European level remain almost constant, temperature increases alter the mixes at country-level, especially the southern countries favour more solar PV. For instance, the wind/solar PV mix of Italy drops from 0.53 to 0.48. Solar PV is ideal to supply cooling demand during daytime, see Figure \ref{fig:cooling and solar}. Such strong correlations between cooling demand and solar PV generations, also seen in \cite{laine2019meeting}, favours solar PV more as temperature increases, particularly in southern Europe.

\begin{figure}[!b]
	\centering
	\includegraphics[trim=0 0cm 0 0,width=0.95\linewidth,clip=true]{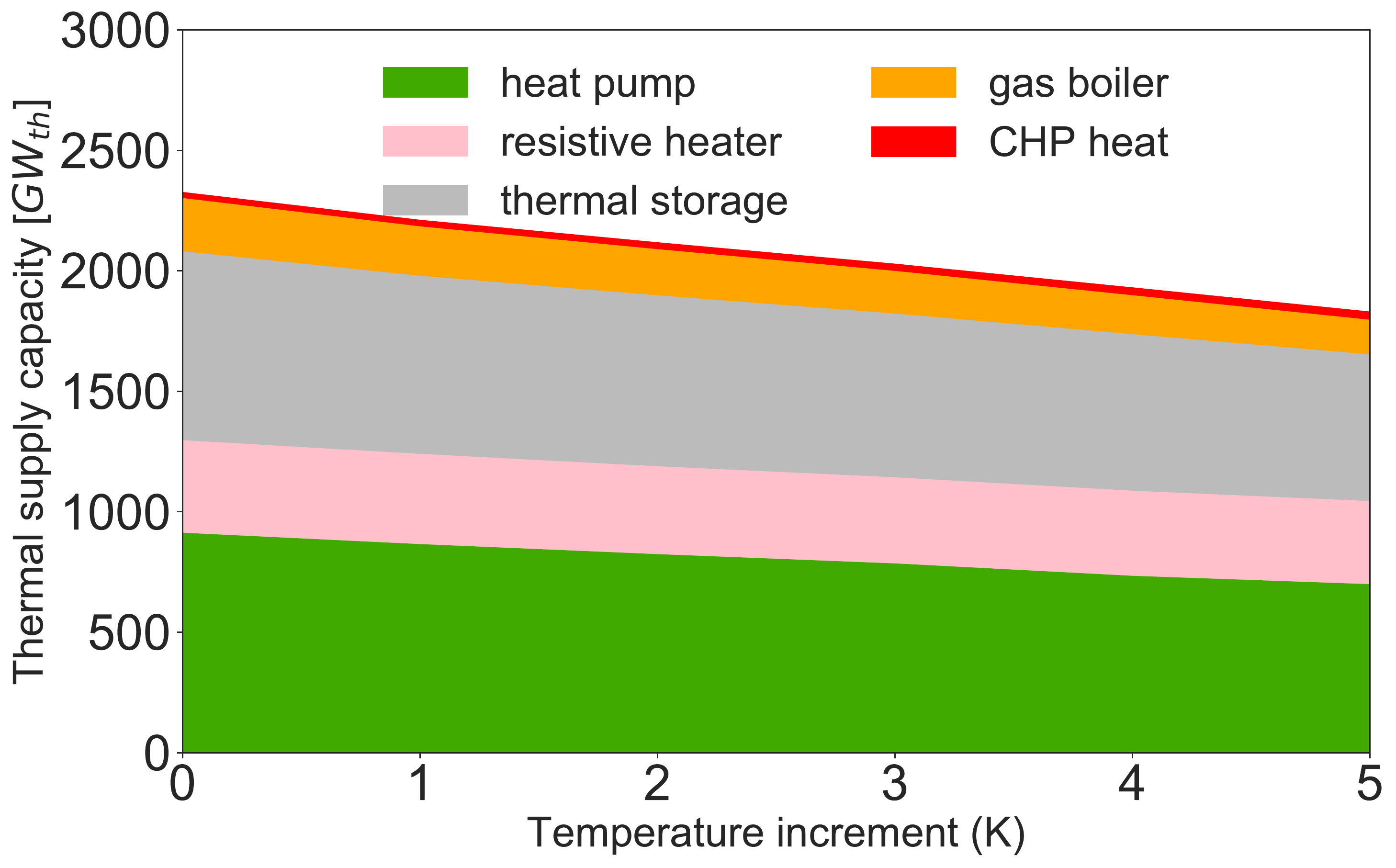}
	\caption{Thermal supply power capacity compositions for temperature increases up to 5\si{\kelvin}. Temperature increase of 0~\si{\kelvin} is equivalent to baseline.}
	\label{fig:TI thermal capacity}
\end{figure} 

As expected, temperature increases have stronger effects on the thermal units, whose supply capacities undergo a substantial reduction, see Figure \ref{fig:TI thermal capacity}. Among thermal units, capacity of gas boilers drops the most. A reduction of more than 25\% is seen for the temperature increase of 5~\si{\kelvin} compared to baseline. Since synthetic gas is used to bridge the gaps in extreme situations, less capacity is required as those situations are smoothed out. Followed by synthetic gas, thermal capacity supplied by heat pumps goes down by a quarter. Capacity of resistive heaters, however, only decrease by 10\% due to its low capital cost and flexibility.

\begin{figure}[!t]
\centering
\includegraphics[trim=0 0cm 0 0,width=\picsize\linewidth,clip=true]{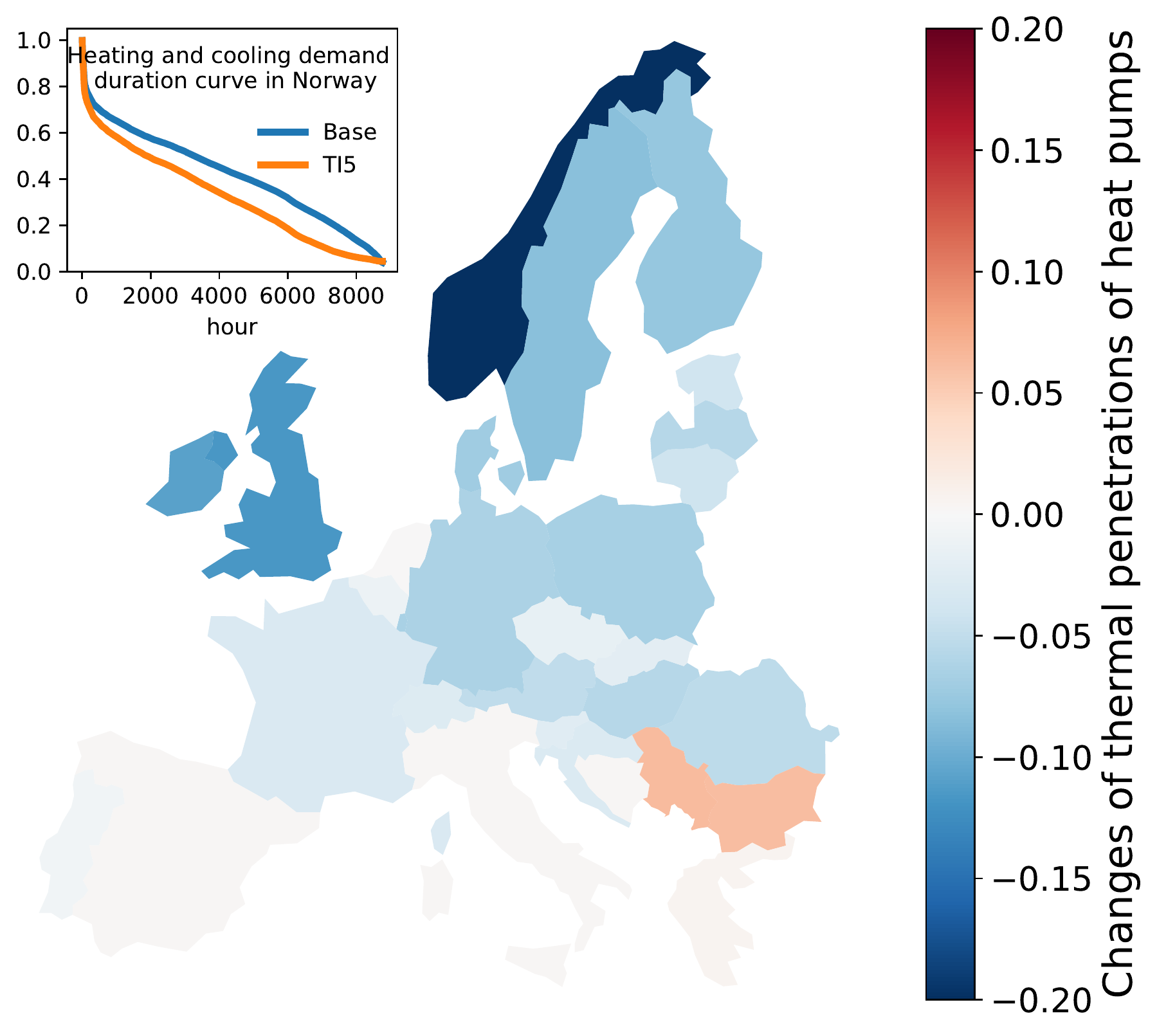}
\caption{Changes of thermal penetrations of heat pumps comparing temperature increase of 5~\si{\kelvin} (labelled by `TI5') to baseline (labelled by `Base'). The inset shows normalised demand duration curve for Norway.}
\label{fig:spatial HP TI5}
\end{figure}

Figure \ref{fig:spatial HP TI5} exhibits the changes of thermal penetrations of heat pumps comparing temperature increase of 5~\si{\kelvin} to baseline on the country level. Most northern countries see negative changes as a result of temperature increase. For instance, thermal penetrations of heat pump in Norway, whose HCD duration curve is depicted in the inset of Figure \ref{fig:spatial HP TI5}, decrease from 68\% to 45\%. The decrement is mainly due to the fact that much lower values of HCD are observed at the tail of Norway's HCD duration curve for temperature increase of 5~\si{\kelvin} compared to baseline. By contrast, a few southern countries have almost the same thermal penetrations or even higher for heat pumps. The inset in Figure \ref{fig:spatial HP Base} shows the HCD duration curve in Spain, where a similar tail is observed for temperature increase of 5~\si{\kelvin} compared to baseline. Even though the total demand of heating and cooling still drops for southern countries, temperature increases result in higher cooling demand, hence requiring more heat pumps to supply it.

Thermal penetrations of resistive heaters increase among most of European countries, and the increments are more pronounced in the north. Heat pumps not only decrease their thermal penetrations, but also their Europe-aggregated utilisation factor, which decreases from 0.62 for baseline to 0.55 for temperature increase of 5~\si{\kelvin}. The lower use of heat pumps for higher temperature scenario can be explained by the strong coupling among various sectors. In the coupled system, surplus electricity can be converted into heating and cooling instead of being curtailed. As total heating and cooling demand decreases but electricity demand remains constant, relatively more surplus electricity is available for power-to-heat units. Therefore, the combination of using more electricity and low-efficient resistive heaters prevails over high-efficient heat pumps.

\subsection{Sensitivity to heat savings}
\begin{figure}[!t]
	\centering
	\includegraphics[trim=0 0cm 0 0,width=\picsize\linewidth,clip=true]{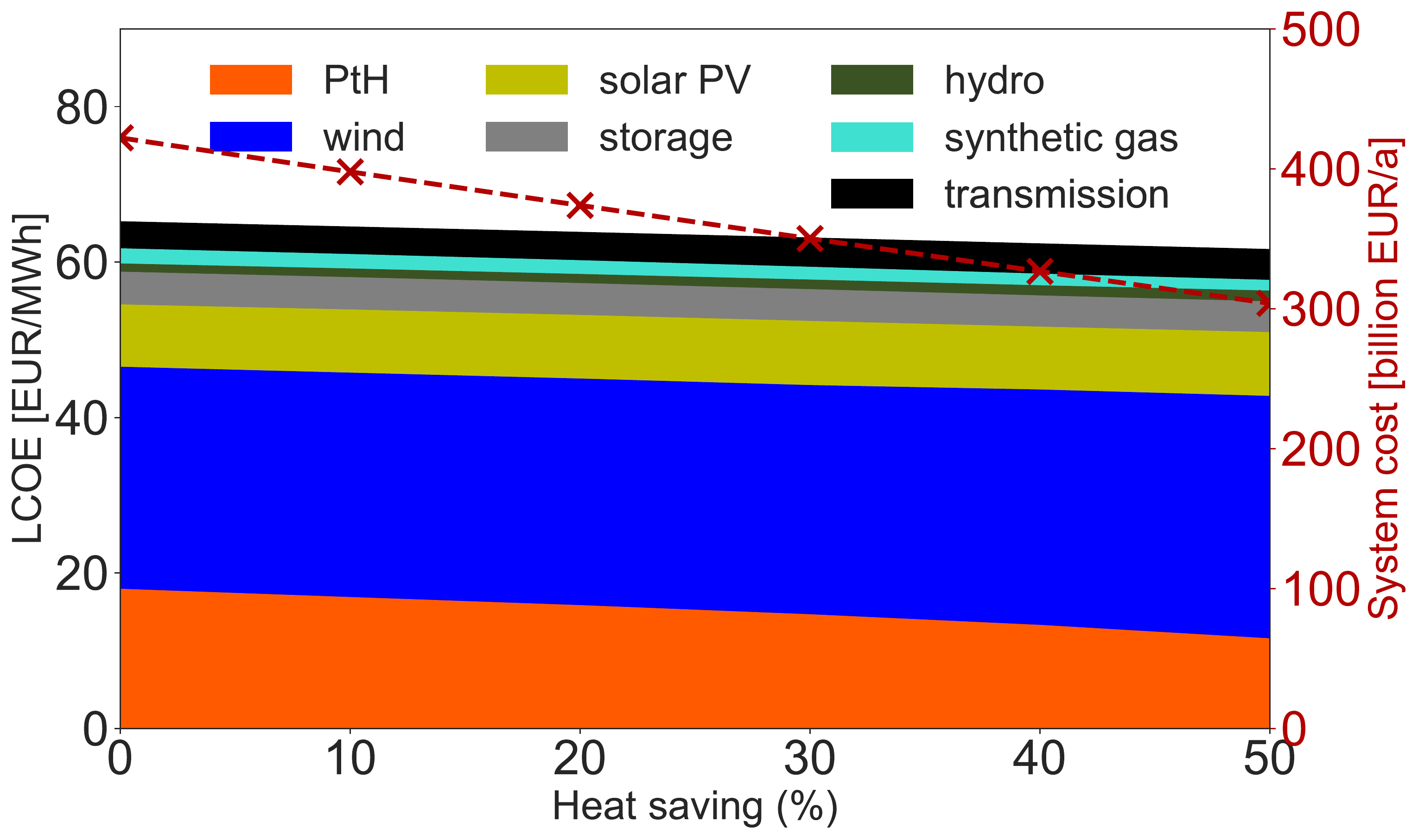}
	\caption{LCOE compositions for heat savings up to 50\%. The red dashed line indicates the annualised system costs. Heat saving of 0\% is equivalent to baseline.}
	\label{fig:HS system cost}
\end{figure}
Heat savings reduce both heating and cooling demand. The sum of LCOE decrease from 65 down to 62~\EUR/MWh for 50\% heat saving, and the decrement is mainly due to lower expenditure of power-to-heat technologies, see Figure \ref{fig:HS system cost}. Lower LCOE and diminished total demand result in a substantial decline in system cost, shown by the red dashed line. Note that the expenses to implement the heat savings are not included in the model, which could potentially raise the LCOE and system costs. 

\begin{figure}[b!]
	\centering
	\includegraphics[trim=0 0cm 0 0,width=0.95\linewidth,clip=true]{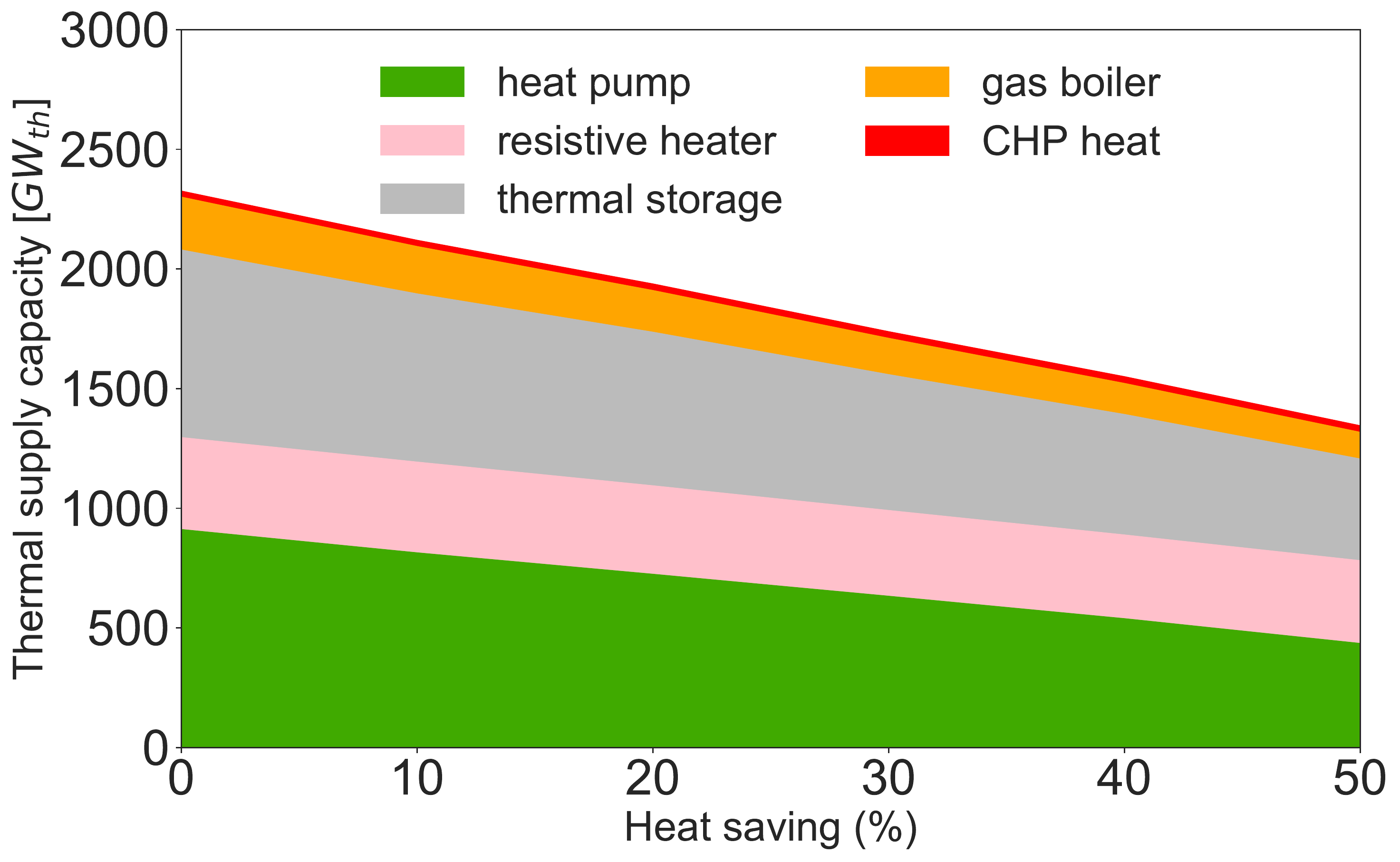}
	\caption{Thermal supply power capacity compositions for heat savings up to 50\%. Heat saving of 0\% is equivalent to baseline.}
	\label{fig:HS thermal capacity}
\end{figure}

Heat savings lead to higher VRES penetrations, but the wind/solar PV mixes remain almost constant. Comparatively, heat savings have stronger impact on the heating and cooling sector. The sum of thermal supply capacities drop significantly, at a slope of 9\% per 10\% of heat saving, see Figure \ref{fig:HS thermal capacity}. Compared to other thermal technologies, resistive heaters show lower decrease, only at a slope of 2\%. In contrast to temperature increases, heat pumps operate at higher utilisation factors due to the homogeneous reduction in heating demand. Apart from that, heat savings lead towards a similar picture compared to temperature increase. Both scenarios lower the HCD, resulting in significantly less heat pumps, but not affecting capacity of resistive heaters. The VRES layouts are only affected to a marginal extent.

\begin{figure}[t!]
	\centering
	\includegraphics[trim=0 0cm 0 0,width=\picsize\linewidth,clip=true]{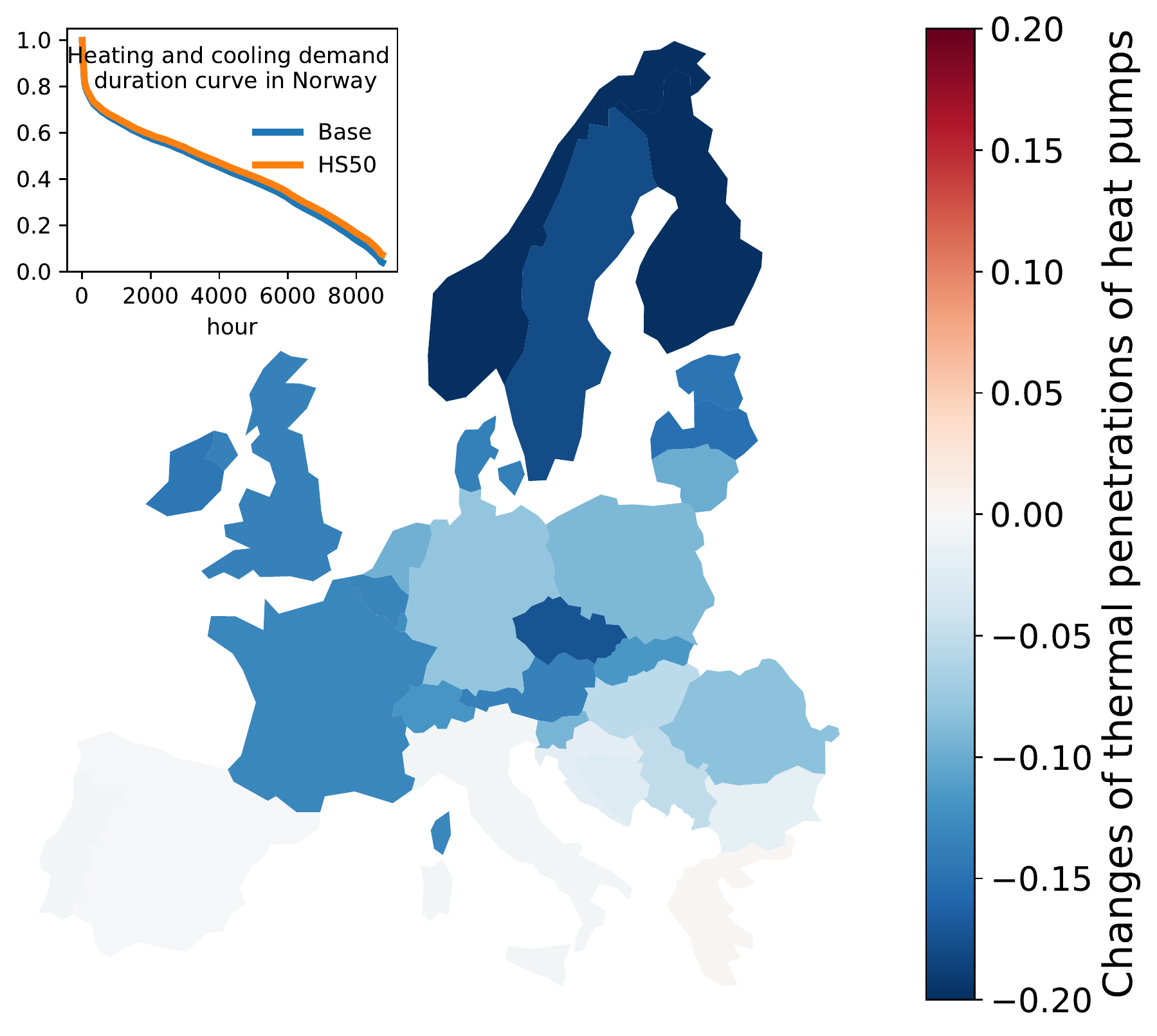}
	\caption{Changes of thermal penetrations of heat pumps comparing heat saving of 50\% (labelled by `HS50') to baseline (labelled by `Base'). The inset shows the normalised demand duration curve for Norway.}
	\label{fig:spatial HP HS50}
\end{figure}

Figure \ref{fig:spatial HP HS50} displays the changes of thermal penetrations for heat pumps comparing heat saving of 50\% to baseline. Despite HCD undergoing a spatially and temporally uniform reduction, most countries end up with lower thermal penetrations of heat pumps, particularly in the northern Europe. For example, thermal penetrations of heat pumps in Norway decrease from 0.68 for baseline to 0.43 for heat saving of 50\%. Even though the HCD duration curve does not significantly change due to heat savings, see the inset of Figure \ref{fig:spatial HP HS50}, lower heating and cooling demand makes surplus electricity more abundant, hence favouring resistive heaters.

\subsection{Sensitivity to demand-side management}
\begin{figure}[t!]
\centering
\includegraphics[trim=0 0cm 0 0,width=\picsize\linewidth,clip=true]{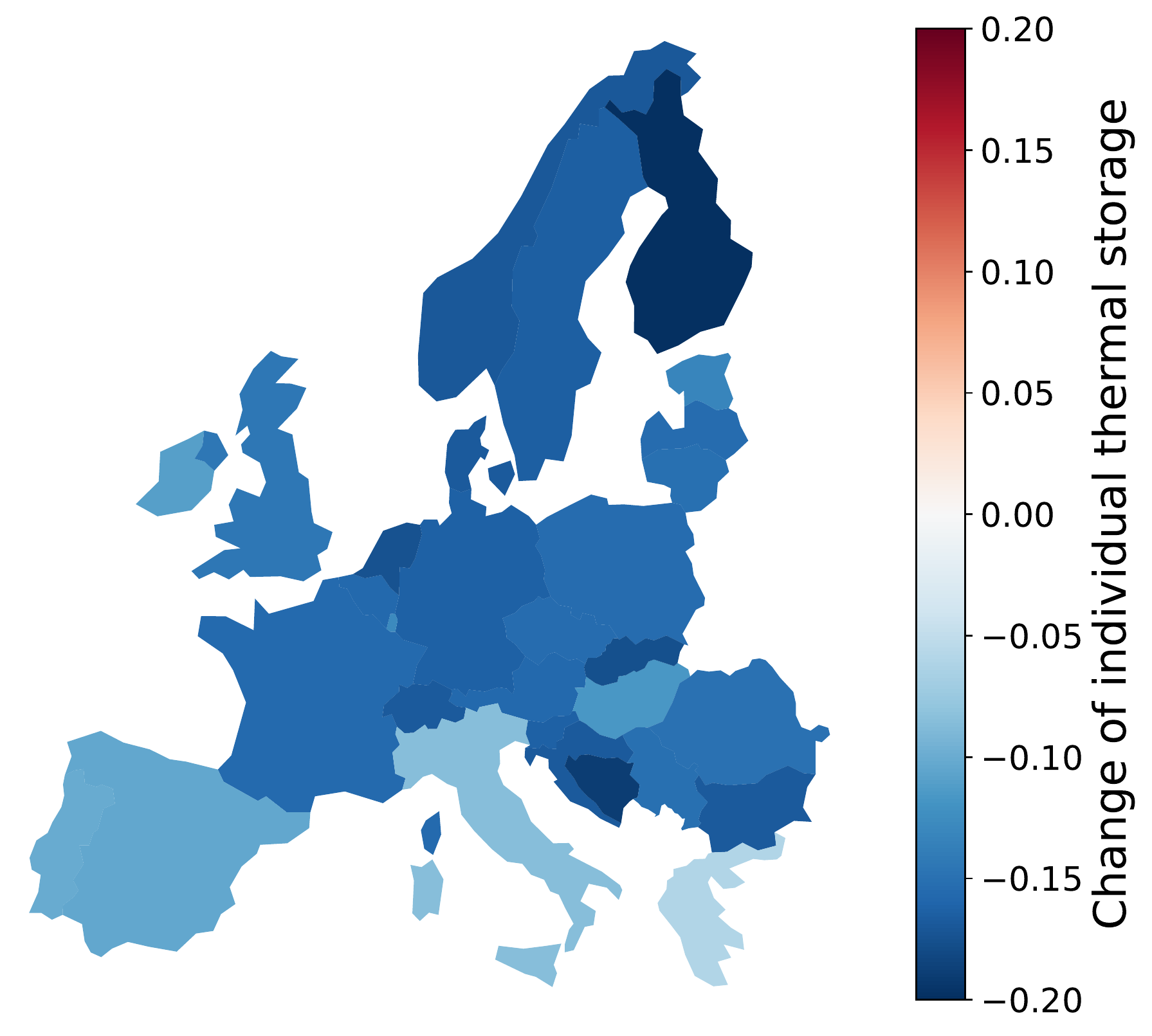}
\caption{Changes of energy capacities for individual thermal energy storage, utilising 6 hours of buildings inertia (labelled by `DM6') compared to baseline.}
\label{fig:DSM}
\end{figure}

Few differences have been found in terms of system costs and configurations for these scenarios. Nevertheless, demand-side management utilising buildings inertia reduces the need for thermal energy storage, particularly for rural areas where seasonal thermal energy storage is not available. Figure \ref{fig:DSM} shows the changes of energy capacities for short-term thermal energy storage, \textit{i.e.}, individual hot water tanks, comparing the scenario of 6 hours demand-side management to baseline. The observed reductions are more pronounced in northern Europe where peak shaping is more effective due to higher diurnal variations of HCD. The benefits of utilising buildings inertia are compromised due to several reasons. First, it is assumed that the storage capacity is limited as over-heating hours and switch-off of the heaters can not exceed a certain period. Second, compared to cheaper and better-insulated hot water tanks, the extra heating stored in the structure thermal mass has higher heat losses. Nevertheless, this way of peak demand shaping could be potentially efficient for the buildings where external thermal energy storage is unavailable or too expensive.

\subsection{Sensitivity to thermal technology cost reductions}\label{sec:TTC}
\begin{figure*}[h!]
	\centering
	\includegraphics[trim=0 0cm 0 0,width=\picsize\linewidth,clip=true]{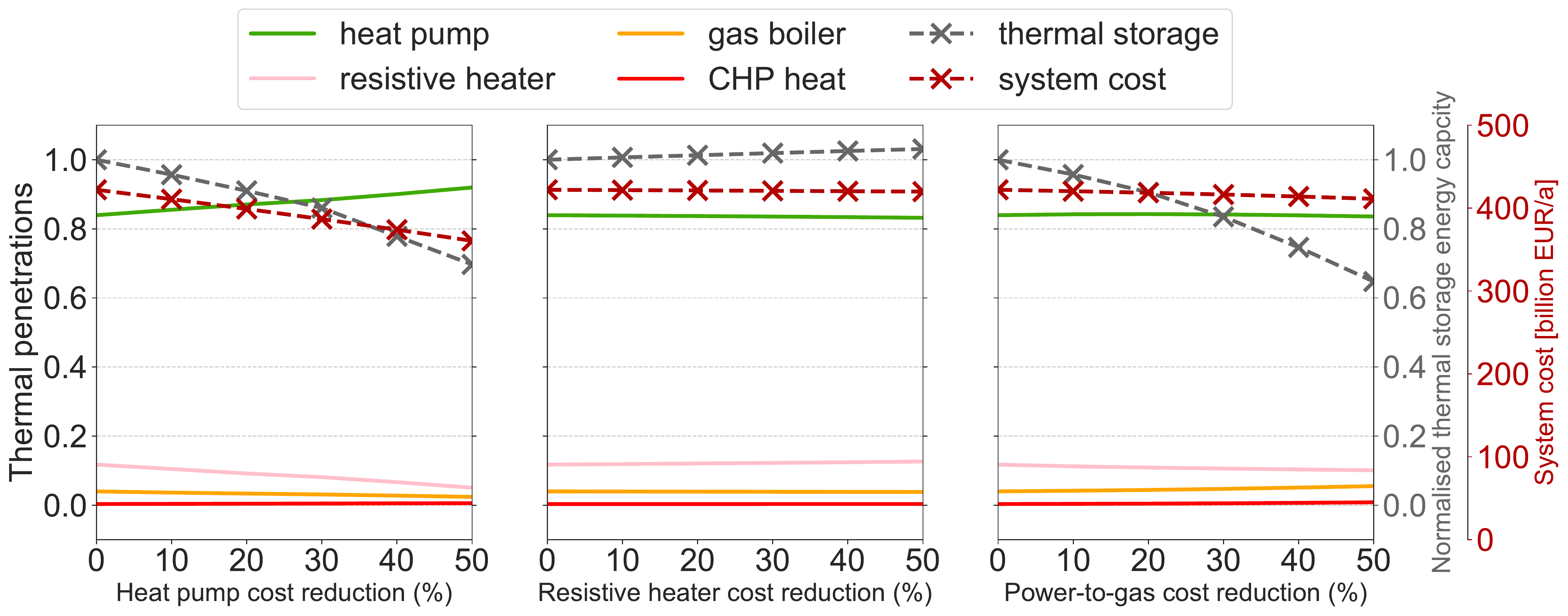}
	\caption{Impact of cost reductions in heat pumps (left), resistive heaters (middle) and power-to-gas (right) on thermal penetrations (left axis), normalised to baseline. Figure also shows thermal energy storage energy capacities (first right axis) and system costs (second right axis). Cost reduction of 0\% is equivalent to baseline.}
	\label{fig:TTC}
\end{figure*}
We investigate here the sensitivity to cost assumptions for different technologies related to heating. As the costs of heat pumps go down, the system costs decline significantly, see the left plot of Figure \ref{fig:TTC}, since heat pumps account for more than 80\% of total heating and cooling generation. Most of decrement comes from lower heat pumps expenditure, while a small extent is contributed by diminishing wind and solar PV installations, since the coupled system requires less electricity because there are more available heat pumps which can efficiently convert electricity into heating. Consequently, the thermal penetrations of heat pumps increase as cost goes down, reaching more than 90\% for 50\% cost reduction. Resistive heaters see a substantial drop in terms of supply capacities and thermal penetrations, while dispatchable backup is barely affected. Since more heating can be supplied by heat pumps directly, the need for thermal energy storage decreases considerably, up to 30\% reduction compared to baseline. Cheaper heat pumps have huge impact on resistive heaters as well as thermal energy storage, but not as much to gas boilers and CHP heat, which are still needed as dispatchable energy. 

System costs barely decrease due to cost reductions of resistive heaters, see the middle plot of Figure \ref{fig:TTC}. Half-priced resistive heaters only save less than 1\% of annual system cost. Despite the thermal supply capacities of resistive heaters increase considerably, up to 25\% comparing 50\% cost reduction to baseline, their thermal penetrations only rise from 11\% to 12\%. Cost reductions do not enable resistive heaters to be cost-efficient compared to heat pumps, since the latter ones have much higher efficiency.

As power-to-gas costs go down, installation of solar PV increases while wind decreases, resulting in a slight decline of system costs, see the right plot of Figure \ref{fig:TTC}. The primary energy generated by wind (solar PV) has changed from 3506 (963)~TWh\el{} for baseline to 3384 (1173)~TWh\el{} for 50\% cost reduction of power-to-gas. The main reason is that large amount of storage is needed to fully utilise solar PV generation due to its high diurnal variations, hence cheaper power-to-gas technologies enable storing surplus solar PV into synthetic gas cost-efficiently. 

Thanks to cheaper power-to-gas technologies, capacities of heat pumps and resistive heaters decrease. Since dispatchable energy becomes cheaper, thermal penetrations of gas boilers increase from 4\% for baseline to 6\% for 50\% cost reduction, and the corresponding decrement comes from resistive heaters. It is also observed that heat losses due to thermal energy storage decrease substantially, from 428 for baseline to 323~TWh\th. Cost reductions of power-to-gas technologies impact resistive heaters and thermal energy storage to a larger extent compared to heat pumps.

\section{Discussions} \label{sec:discussions}
The baseline scenario introduced in this paper incorporates net-zero emissions for the electricity, heating and cooling coupled European energy system. The annual system cost is 422~billion~\EUR and LCOE is 65~\EUR/MWh. Those values are respectively 19\% and 17\% higher than the scenario of 95\% \co{} reduction discussed in \cite{zhu2019impact}. VRES and PtH capacities account for most of the increment, while the expenditure for gas drop significantly due to stricter emissions constraint. The last 5\% of emissions drives up the system cost, also seen in \cite{brown2019sectoral}, indicating the difficulty of achieving fossil-free system, yet feasible.

Kozarcanin \textit{et al.} \cite{kozarcanin2019impact} find heat pumps to become more competitive with the increasing ambient temperature of the different climate change scenarios. They identify two main causes for this. First, increasing ambient temperature enhance the COP of heat pumps. Second, the utilisation factors of the selected heating technology increase as temperature increases. We find here an opposite result: heat pump optimal thermal penetrations decrease with increasing temperatures. The reason for this discrepancy is related to sector coupling. While Kozarcanin \textit{et al.} identify the cost-optimal heating technologies assuming an independent heating sector, we co-optimize the coupled electricity, heating and cooling system. Climate change scenarios result in lower heating demand, while electricity demand remains constant. Therefore, it is cost-optimal to increase the capacity of resistive heaters to convert the more abundant surplus electricity into heating that can be instantaneously consumed or stored.

The impact of climatic, technical and economic uncertainties are analysed on top of the baseline scenario. The methodology employed in this paper has some limitations which are described below. To investigate the impact of climate change, a simplified approach with homogeneous temperature increases has been introduced in this paper. Similarly, assuming a uniform building retrofitting for the whole Europe may overlook the spatial characteristics among different countries. Demand-side management strategy utilising building structure thermal mass has been modelled on top of thermal energy storage in this paper, whereas other forms of demand-side management, such as demand response, load shedding, are worthy of further investigations. The uncertainties caused by cost assumptions are investigated in three categories, \textit{i.e.}, heat pumps, resistive heaters and power-to-gas technologies, while a same reduction rate has been applied on different technologies within one category. It would be interesting to adopt a more comprehensive sensitivity analysis on the cost assumptions to further understand the synergy among various technologies.

Our model assumes perfect foresight and long-term market equilibrium, and it minimises the system cost including capacity and dispatch expenditures, in order to maximise social welfare. Using different approaches, \textit{e.g.}, synchronised dispatch scheme \cite{eriksen2017optimal} or rule based \cite{connolly2014heat}, would very likely have an inevitable impact on the results drawn in Section \ref{sec:results}. Likewise, estimating wind/solar PV capacity factors or geographical potentials in different manners might lead to distinct optimal VRES configurations throughout Europe. 

To limit temperature increment below 1.5\si{\celsius} by the end of the century, European countries demand net-zero \co{} emissions for the whole energy sector. Coupling to additional sectors, particularly transport and industry, brings extra challenges, but also provides more flexibility and might reveal extra synergies among different sectors. For instance, a highly electrified transport sector could eliminate the need for costly static electricity storage \cite{VICTORIA2019111977}, or electro-fuels could be used to couple the electricity and transport sector \cite{mathiesen2015smart}. Potential demand-side management of charging electric vehicles could facilitate smoothing of the electricity load on the other hand. In addition, our model does not include the gas grid which could also be used for transporting hydrogen to reinforce its role in the system \cite{welder2018spatio,welder2019design,blanco2018potential}.

A more comprehensive selection of technologies, in particular renewable resources such as biofuels, solar thermal, geothermal, or concentrated solar power, could facilitate to decarbonise the energy system. Extensive choices of mature low-carbon technologies could offer additional flexibilities to the existing model, hence bringing system costs further down. It would also be interesting to include alternative non-renewable technologies, like nuclear or carbon capture and storage, and find out their roles in deeply decarbonised energy systems. 

A coarse-grained network is implemented for the model, where each country is aggregated into one single node. This simplification may neglect transmission bottlenecks, as well as spatial variations of VRES within individual country. Moreover, the simulations are carried out for only one year, where the demand and VRES capacity factors time series corresponding to 2015 are used. 

\section{Conclusions} \label{sec:conclusions}
To evaluate the impact of climatic, technical and economic uncertainties on the design of a coupled electricity, heating and cooling system in Europe, an hourly-resolved, one-node-per-country network model with net-zero \co{} emissions constraint has been used in this paper. On top of it, scenarios including synthetic temperature increases, heating savings, demand-side management and cost reductions of key components are applied to quantify the impact of uncertainties.

For the baseline scenario, system cost is dominated by wind, solar PV and power-to-heat capacities, while the remainder is spent on balancing the renewable generation spatially by transmission and temporally through storage. Heat pumps provide most of heating and cooling demand at a rather constant output thanks to high efficiency. Urban areas rely heavily on long-term thermal energy storage, while rural areas consume synthetic gas through boilers. The synergy between long-term storage provided by synthetic gas and large hot water tanks, inexpensive resistive heaters and high-efficient heat pumps manage to carry out a coupled fossil-free electricity, heating and cooling system.

Scenarios with synthetic climate change entail reduced annual heating demand and increased cooling demand, but the impact of the former is higher. Lower heating demand reduces the total system cost but has a negligible impact on the LCOE. The Europe-averaged optimal wind/solar mix remains roughly unchanged, but southern countries install higher solar PV capacities to benefit from the strong temporal correlation between solar generation and cooling demand. With a reduced heating demand, more surplus electricity is available for power-to-heat technologies, which results in increment of the optimal thermal penetrations of resistive heaters at the expense of heat pumps. The low installation cost of the former does not penalize lower utilization factors and, hence, they can be used to transform otherwise-curtailed electricity into heat.

Heat saving scenarios assume a spatially and temporal homogeneous reduction for heat and cooling demand, which result in lower system cost and LCOE. Although the cost savings need to be compared to the cost of building retrofitting, this result highlights the potential of reducing demand to facilitate the design and operation of the energy system. The optimal thermal penetrations of heat pumps also diminish due to the reasons previously mentioned. 

We find that including demand-side management strategies, here modelled by using building structure thermal mass to allow storing heat up to ten hours, do not modify the optimal system cost or configuration. The reason behind this is that large hot water tanks installed in centralised heating systems and individual tanks in rural areas already provide an effective way of smoothing heating demand.

The sensitivity to cost assumptions for heating technologies is also evaluated. As heat pumps supply most of the heating demand in the baseline scenario, decreasing their cost has the most significant impact, reducing system cost and increasing the optimal thermal penetration of this technology. Resistive heaters and power-to-gas represent less than 5\% of the system cost. Hence, a cost reduction of any of them modifies neither the system cost nor the share of heat supply. However, since synthetic gas serves as storage, cost reduction of power-to-gas translates into a lower need for thermal energy storage.

In this paper, we focus on the coupled electricity, heating and cooling European system and show how different uncertainties could impact the optimal system configuration. The analysis carried out here should be expanded by including additional sectors such as industry and transport to exploit all the benefits of sector coupling. It will also be interesting to address the sensitivity to climate, technical and economic assumptions of such a further-integrated system.

\section*{Acknowledgements}
K. Zhu, M. Victoria, M. Greiner and G. B. Andresen are fully or partially funded by the RE-INVEST project, which is supported by the Innovation Fund Denmark under grant number 6154-00022B. The responsibility for the contents lies solely with the authors. 

\section*{References}
\bibliography{mybibfile}

\appendix
\section{PyPSA-Eur-Sec-30 Model} \label{sec:appendix}
The model is implemented as a techno-economical optimisation problem, which minimises the total system costs expressed as a linear function (Eq.\ (\ref{eq:objective})) subject to technical and physical constraints (Eqs.\ (\ref{eq:energybalance}) - (\ref{eq:alpha})), assuming perfect competition and foresight. The open-source framework PyPSA \cite{brown2017pypsa} and the PyPSA-Eur-Sec-30 model introduced in \cite{brown2018synergies}, are used. Each of the 30 European countries covered by the model is aggregated into one node, which consists of one electricity bus, two heat buses for urban and rural areass, and one cooling bus (see in Figure \ref{fig:flow}). Neighbouring countries are connected through cross-border transmission lines, including existing and under construction lines (see Figure \ref{fig:topology and demand}). High Voltage Direct Current (HVDC) is assumed for the transmission lines, whose capacities can be expanded by the model if it is cost-effective. Within each country, different buses are connected by energy converters as shown in Figure \ref{fig:flow}.

The model runs over a full year of hourly data. The inelastic loads of electricity, heating which includes the ratio between urban and rural heating, and cooling, are exogenous to the model and not optimised, as well as for hydroelectricity, \textit{i.e.}, hydro reservoir, run-of-river generators, and pumped-hydro storage, where fixed capacities are assumed due to environmental concerns. By contrast, VRES generator capacities, \textit{i.e.}, onshore wind, offshore wind, and solar PV, conventional generator capacities, \textit{i.e.}, open cycle gas turbines (OCGT), combined heat and power (CHP), gas boilers, converter capacities, \textit{i.e.}, heat pumps and resistive heaters, storage power and energy capacities, \textit{i.e.}, batteries and hydrogen for electricity and hot water tanks for heating, and transmission capacities are all optimised. In addition, the hourly operational dispatch of generators, converters, and storage units are subject to optimisation as well.

\subsection{Objective function}
As mentioned in Section \ref{sec:methods}, each country $i$ has four buses labelled by $n$. Generators and storage technologies are denoted by $s$, hour of the year by $t$, and bus connectors by $\ell$, which include both transmission lines and converters. The total annual system cost consists of fixed annualised costs $c_{n,s}$ for generator and storage power capacity $G_{n,s}$, fixed annualised costs $\hat{c}_{n,s}$ for storage energy capacity $E_{n,s}$, fixed annualised costs $c_\ell$ for bus connectors $F_{\ell}$, variable costs, for generation and storage dispatch $g_{n,s,t}$. The total annual system cost is minimised by:
\begin{align}
& \min_{\substack{G_{n,s},E_{n,s},\\F_\ell,g_{n,s,t}}} \left[ \sum_{n,s} c_{n,s} \cdot G_{n,s} +\sum_{n,s} \hat{c}_{n,s} \cdot E_{n,s} \right. \nonumber \\
& \hspace{2cm} \left. + \sum_{\ell} c_{\ell} \cdot F_{\ell}+ \sum_{n,s,t} o_{n,s,t} \cdot g_{n,s,t} \right]
\label{eq:objective}
\end{align}

\subsection{Constraints}
The demand $d_{n,t}$ of bus $n$ at hour $t$ is met by VRES generation, hydroelectricity, conventional backup (OCGT, CHP, gas boiler), storage discharge, converters (heat pump, resistive heater) and HVDC transmission across border.
\begin{equation}
\sum_{s} g_{n,s,t}+ \sum_{\ell} \alpha_{n,\ell,t}\cdot f_{\ell,t} = d_{n,t} \hspace{.2cm} \leftrightarrow \hspace{0.2cm} \lambda_{n,t} \hspace{.3cm} \forall\, n,t \label{eq:energybalance}
\end{equation}
where $f_{\ell,t}$ refers to energy flow on the link $l$ and $\alpha_{n,\ell,t}$ indicates both the direction and the efficiency of flow on the bus connectors; it can be time-dependent such as heat pumps. The Lagrange/Karush-Kuhn-Tucker (KKT) multiplier $\lambda_{n,t}$ associated with the demand constraint represents the local marginal price of the energy carrier. 

The dispatch of generators and storage is bounded by the product between installed capacity $G_{n,s}$ and availabilities $\ubar{g}_{n,s,t}$, $\bar{g}_{n,s,t}$:
\begin{equation}
\ubar{g}_{n,s,t} \cdot G_{n,s} \leq g_{n,s,t} \leq \bar{g}_{n,s,t} \cdot G_{n,s} \hspace{1cm} \forall\, n,s,t \;  \label{eq:gen}
\end{equation}
$\ubar{g}_{n,s,t}$ and $\bar{g}_{n,s,t}$ are time-dependent lower and upper bounds due to, \textit{e.g.}, VRES weather-dependent availability. For instance, for wind generators, $\ubar{g}_{n,s,t}$ is zero and $\bar{g}_{n,s,t}$ refers to the capacity factor at time $t$. $G_{n,s}$ is the installed power capacity for generators, limited by installable potentials $\bar{G}_{n,s}$ due to, \textit{e.g.}, geographical constraints:
\begin{equation}\label{eq:geo limit}
0 \leq G_{n,s}\leq \bar{G}_{n,s} \hspace{1cm} \forall\, n,s \; 
\end{equation}
Similarly, the dispatch of converters has to fulfil the following constraints
\begin{equation}
\ubar{f}_{\ell,t} \cdot F_{\ell} \leq f_{\ell,t} \leq \bar{f}_{\ell,t} \cdot F_{\ell} \hspace{1cm} \forall\, \ell,t \;  \label{eq:con}
\end{equation}
For a unidirectional converter, \textit{e.g.}, a heat pump, $\ubar{f}_{\ell,t}=0$ and $\bar{f}_{\ell,t}=1$ since a heat pump can only convert electricity into heating. For transmission links, $\ubar{f}_{\ell,t}=-1$ and $\bar{f}_{\ell,t}=1$, which allows both import and export between neighbouring countries. In particular, the inter-connecting transmission can be limited by a global constraint
\begin{equation}
\sum_{\ell} l_\ell \cdot F_{\ell} \leq  \textrm{CAP}_{LV} \hspace{.7cm} \leftrightarrow \hspace{0.3cm} \mu_{LV} \; 
\label{eq:lvcap}
\end{equation}
where the sum of transmission capacities $F_{\ell}$ multiplied by the lengths $l_{\ell}$ is bounded by a transmission volume cap $\textrm{CAP}_{LV}$. The KKT multiplier $\mu_{LV}$ associated with the transmission volume constraint indicates the shadow price of an increase in transmission volume to the system. 

The state of charge $e_{n,s,t}$ of every storage has to be consistent with charging and discharging in each hour, and is limited by the energy capacity of the storage $E_{n,s}$
\begin{align}
e_{n,s,t} = & \ \eta_0 \cdot e_{n,s,t-1} + \eta_{1} |g_{n,s,t}^+| - \eta_{2}^{-1} |g_{n,s,t}^-| \nonumber \\
& + g_{n,s,t,\textrm{inflow}} - g_{n,s,t,\textrm{spillage}} \; , \nonumber \\
& 0  \leq   e_{n,s,t} \leq E_{n,s}   \hspace{0.5cm} \forall\, n,s,t \; . \label{eq:soc}
\end{align}
The storage has a standing loss $\eta_0$, a charging efficiency $\eta_1$ and rate $g_{n,s,t}^+$, a discharging efficiency $\eta_2$ and rate $g_{n,s,t}^-$, possible inflow and spillage which are subject to (Equation \ref{eq:gen}). The storage energy capacity $E_{n,s}$ can be optimised independently of the storage power capacity $G_{n,s}$.

To enforce the decarbonisation of the energy system, a net-zero emissions constraint imposed to the model, which means only synthetic gas is allowed to be consumed as dispatchable backup for OCGT, CHP and gas boiler, and the sum of initial filling level ($e_{n,s,t=0}$) for gas storage has to be equal to the ending level ($e_{n,s,t=T}$)
\begin{equation}
\sum_{n,s}(e_{n,s,t=0}-e_{n,s,t=T}) = 0 \hspace{.5cm} \leftrightarrow \hspace{0.3cm} \mu_{CO_2} \label{eq:CO2 price}
\end{equation}
The KKT multiplier $\mu_{CO_2}$ indicates the necessary carbon emissions tax to fulfil this constraint in an open market.

\subsection{VRES layout} \label{sec:VRES_layout}
For every country $i$, the annual available VRES generation is denoted by $g_{i,VRES}$, representing the energy that can be potentially generated, that is, before curtailment.
\begin{equation*}
g_{i,VRES} = \sum_{t,s\in VRES,n\in i} \bar{g}_{n,s,t} \cdot G_{n,s}
\end{equation*}
The VRES penetration $\gamma_i$ is defined as the ratio of VRES generation to the total demand in country $i$, which is the sum of electricity, heating and cooling demand
\begin{equation}
g_{i,VRES} = \gamma_i \sum_{t,n\in i} d_{n,t} \label{eq:gamma}
\end{equation}
The VRES generation consists of wind generation $g_{i,W}$ and solar $g_{i,S}$,
\begin{equation*}
g_{i,VRES} = g_{i,W}+g_{i,S} \; 
\end{equation*}
The wind/solar mix parameter $\alpha_i$ determines the ratio between available wind and VRES
\begin{equation*}
g_{i,W} = \alpha_i \cdot g_{i,VRES} \; 
\end{equation*}
The VRES mix of the whole system $\alpha_{EU}$ can be found by
\begin{equation}
\sum_{i} g_{i,W} = \alpha_{EU} \cdot \sum_{i} g_{i,VRES} \label{eq:alpha EU}
\end{equation}
where $\alpha_{EU}$ expresses the overall VRES layout tendency towards wind or solar dominance. 

To utilise different VRES resources over the continent, a `weakly homogeneous' layout \cite{eriksen2017optimal} is introduced. `Homogeneous' makes sure that the share of VRES in each country compared to its total demand is the same, hence $\gamma_i$ can be shortened to $\gamma$. `Weakly' suggests that the wind/solar PV mix $\alpha_i$ of each country is optimised.
\begin{equation}
\gamma_i = \gamma, \hspace{0.2cm} \alpha_i \textrm{ subject to opt} \label{eq:alpha}
\end{equation}
This layout ensures that each country is VRES self-sufficient to a certain extent, and the optimisation seeks the optimal wind/solar mix in each country.

\end{document}